\documentclass[12pt]{article}
\pdfoutput=1
\usepackage{graphicx}
\usepackage{epstopdf}
\usepackage{amsmath}
\usepackage{amsfonts}
\usepackage{amssymb}
\usepackage{color}
\usepackage{mathrsfs}

\usepackage[font={footnotesize}]{caption}

\pdfoutput=1

\usepackage[english]{babel}

\usepackage[colorlinks,bookmarks]{hyperref}
\definecolor{linkblue}{rgb}{0,0,0.8}
\definecolor{linkgreen}{rgb}{0,0.5,0}

\hypersetup{pdfpagemode=UseNone, pdfstartview=FitH, linkcolor=linkblue, %
            citecolor=linkgreen, urlcolor=linkblue}

\numberwithin{equation}{section}

\newcommand{\hinvMpc}{\,h\, {\rm Mpc}^{-1}\,}

\newcommand{\bea}{\begin{eqnarray}}
\newcommand{\eea}{\end{eqnarray}}
\newcommand{\be}{\begin{equation}}
\newcommand{\ee}{\end{equation}}
\newcommand{\fr}[2]{\frac{ #1}{#2}}
\newcommand{\n}{\nonumber \\}
\newcommand{\cH}{\mathcal{H}}
\newcommand{\kv}{\vec{k}}
\newcommand{\qv}{\vec{q}}

\newcommand{\la}{\langle}
\newcommand{\ra}{\rangle}
\newcommand{\df}{\delta^{(1)} }
\newcommand{\dfp}{\delta^{(1)'} }
\newcommand{\dfpp}{\delta^{(1)''} }

\newcommand{ \bk }{ \vec{k}}
\newcommand{\bq}{\vec{q}}
\newcommand{\bx}{\vec{x}}
\newcommand{\bv}{\vec{v}}
\newcommand{\knl}{k_{\rm NL}}
\newcommand{\kvec}{\vec{k}}
\newcommand{\qvec}{\vec{q}}
\newcommand{\half}{\frac{1}{2}}

\newcommand{\edl}{\epsilon_{\delta < } }
\newcommand{\esg}{\epsilon_{s >}}
\newcommand{\esgrel}{\epsilon_{s >}^{\rm rel}}
\newcommand{\esl}{\epsilon_{s<}}
\newcommand{\eslrel}{\epsilon_{s<}^{\rm rel}}
\newcommand{\eqn}[1]{Eq.~(\ref{#1})}
\newcommand{\dca}{\Delta \bar{c}_A^2}
\newcommand{\ccs}{\bar{c}^2_{c,w_b=0}}
\newcommand{\ci}{\bar{c}_I^2}
\newcommand{\ca}{\bar{c}_A^2}

\newcommand{\unitscss}{ \, (h { \rm Mpc^{-1}})^{-2} }
\newcommand{\unitsk}{\, h { \rm Mpc^{-1}}}

\newcommand{\pd}{\partial}

\newcommand{\Comment}[1]{{}}

\begin{document}
\def\thefootnote{\fnsymbol{footnote}}


\setcounter{page}{1} \baselineskip=15.5pt \thispagestyle{empty}

\begin{flushright}
\end{flushright}

\begin{center}

{\Large \bf  Analytic Prediction of Baryonic Effects  \\[0.3cm] from the EFT of Large Scale Structures \\[0.7cm]}
{\large Matthew Lewandowski${}^{1,2}$,  Ashley Perko${}^1$, and Leonardo Senatore$^{1,2}$}
\\[0.7cm]
{\normalsize { \sl $^{1}$ Stanford Institute for Theoretical Physics,\\ Stanford University, Stanford, CA 94306}}\\
\vspace{.3cm}

{\normalsize { \sl $^{2}$ Kavli Institute for Particle Astrophysics and Cosmology, \\
Physics Department and SLAC, Menlo Park, CA 94025}}\\
\vspace{.3cm}

\end{center}

\vspace{.8cm}

%
%
%

\hrule \vspace{0.3cm}
{\small  \noindent \textbf{Abstract} \\[0.3cm]
\noindent The large scale structures of the universe will likely be the next leading source of cosmological information. It is therefore crucial to understand their behavior. The Effective Field Theory of Large Scale Structures provides a consistent way to perturbatively predict the clustering of dark matter at large distances. The fact that baryons move distances comparable to dark matter allows us to infer that baryons at large distances can be described in a similar formalism: the backreaction of short-distance non-linearities and of star-formation physics at long distances can be encapsulated in an effective stress tensor, characterized by a few parameters. 
The functional form of baryonic effects can therefore be predicted. In the power spectrum the leading contribution goes as $\propto k^2 P(k)$, with $P(k)$ being the linear power spectrum and with the numerical prefactor depending on the  details of the star-formation physics. We also perform the resummation of the contribution of the long-wavelength displacements, allowing us to consistently predict the effect of the relative motion of baryons and dark matter. We compare our predictions with simulations that contain several implementations of baryonic physics, finding percent agreement up to relatively high wavenumbers such as $k\simeq 0.3\hinvMpc$ or $k\simeq 0.6\hinvMpc$, depending on the order of the calculation. Our results open a novel way to understand baryonic effects analytically, as well as to interface with  simulations.

 \vspace{0.3cm}
\hrule


%
%
\def\thefootnote{\arabic{footnote}}
\setcounter{footnote}{0}

%
%
%
%

%
%
%
%

\section{Introduction and Main Idea}

After the completion of the data analyses of the Planck satellite, the next leading source of cosmological information will likely be large scale structure (LSS) surveys. The cosmological information that we inherited from the WMAP and Planck missions raises the bar extremely high: in order for LSS to be able to significantly improve our knowledge of the early universe, it is mandatory to understand to percent level the behavior of the LSS observables. Order-of-magnitude understanding very rarely will be useful. Since most of the modes are gathered at short distances, this means that we need to understand the quasi-linear regime of structure formation. Recently, a research program called the Effective Field Theory of Large Scale Structures (EFTofLSS) has been launched~\cite{Baumann:2010tm,Carrasco:2012cv,Carrasco:2013mua,Angulo:2014tfa,Baldauf:2014qfa}, with the purpose of developing a consistent approach to analytically studying LSS in the weakly non-linear regime. The approach is based on the following observation. At non-linear level, arbitrarily short modes contribute at long  distances, by the simple fact that the product of two short-wavelength modes $\vec k_S$ and $\vec k_S-\vec k_L$, where $_S$ stays for short, and $_L$ stays for long, give rise to a long mode $\vec k_L$. However, in LSS modes shorter than the so-called non-linear scale, which is about $10\,$Mpc, are not under perturbative control. In the EFTofLSS, the equations of motion for the long wavelength modes contain some terms whose role is to encode at long distances the effect of the modes shorter than the non-linear scale. In the case of dark matter, this manifests itself in the fact that the resulting equations of motion take the form of a fluid-like system, with an effective stress tensor that contains a speed of sound, viscosity, a stochastic term, etc.~\cite{Carrasco:2012cv,Baumann:2010tm}. There parameters cannot be predicted in the theory but need to be measured either {\it directly} in observations or in simulations~\cite{Carrasco:2012cv}. Symmetries dictate the form of these terms, which allows the theory to correctly encode the effect at long distances  of short fluctuations by correcting the wrong contribution that convolution loops generate at long distances when integrating over short modes. This is what is called `renormalization'. When applied to collapsed object, this same  phenomenon shows itself into bias coefficients ~\cite{Senatore:2014eva}. Similarly, when considering redshift space distortions, the change of coordinates that maps real space into redshift space is sensitive to very short distance fluctuations in the density and the velocity fields, which also require the addition of extra parameters to correctly reproduce the effect of short distance physics at long distances~\cite{Senatore:2014vja}~\footnote{One might naively think that by cutting off the convolution loops at modes of order the non-linear scale, the effect of non-linear modes is not included. Such a procedure is incorrect  because even if we cannot reliably compute the properties of non-linear modes, they {\it do} have effect at long distances, and this cannot be neglected for the correct predictions of long distance correlation functions. A proof of this fact is that such a naive procedure would give results which are cutoff dependent, while clearly physical observables are not dependent on the cutoff.}. 

When compared to real space dark matter correlations from numerical simulations, the EFTofLSS has performed remarkably well. At one loop, the matter power spectrum agrees with $N$-body simulations to percent level up to wavenumber $k\sim 0.3\hinvMpc$~\cite{Carrasco:2012cv}. Predictions depend on one free parameter,   the so-called speed of sound, which can be measured either by fitting to the power spectrum or directly in small $N$-body simulations~\cite{Carrasco:2012cv}. At one loop the momentum power spectrum~\cite{Senatore:2014via} and the bispectrum~\cite{Angulo:2014tfa, Baldauf:2014qfa} agree with similar accuracy to $N$-body data up to the same wavenumber $k\sim 0.3\hinvMpc$. These are important consistency checks of the EFTofLSS, as the EFT is expected to perform equally well on all correlation functions when they are computed at the same loop order. Another important consistency check passed by the EFTofLSS has been the prediction of the slope of the velocity vorticity power spectrum which matches the one measured in simulations~\cite{Carrasco:2013mua,Mercolli:2013bsa}. 

At two loops the EFTofLSS agrees with the power spectrum of dark matter up to a very high wavenumber $k\sim 0.6\hinvMpc$~\cite{Carrasco:2013mua,Senatore:2014via}. This is an extremely interesting result because it suggests that we can have perturbative control on very small scales. If the same reach is maintained in all observables, the consequences  for next generation large scale structure surveys could be huge.  The number of available modes would increase by a large factor which would lead to very significant improvements for the capability of these experiments to constrain properties of neutrinos, dark energy, the primordial power spectrum, and, most importantly, primordial non-Gaussianities. 

While we believe that these are quite remarkable results, it is a fact that the physics described by dark matter can be simulated by $N$-body codes, so one might argue what is the point of developing analytics techniques when one could just do simulations. First, we believe it is always useful to have an analytic way to understand the dynamics when this is still simple, as in the quasi-linear regime. Secondly, and most importantly, the analytic techniques from the EFTofLSS are relatively simple, easy to check, and fast: they do not require complicated and computationally expensive numerical codes~\footnote{This has consequences on the actual usability of the results. One of the authorS of this paper has experience with interacting with $N$-body codes, and has seen the difficulty in extracting precise results from them.}. We believe such a simplicity has the potential of allowing us to make much progress in understanding the LSS. In fact, while one might be skeptical of the arguments just mentioned, one can simply look at the comparison of the information we are currently gathering between the CMB and LSS. Most of the data in LSS surveys are not used because of lack of theoretical control. This leads to LSS giving a significant contribution to our knowledge of the universe mainly when they break some degeneracies in the CMB, as in the case of dark energy. The contribution is very little when the CMB is not degenerate. Clearly, in the light of the fact that in the very near future we will likely start to be dominated by LSS surveys, the current level of understanding cannot be considered satisfactory.  After the Planck satellite has completed, if we want to make further progress, the situation has to change.

It should be stressed that in the case of dark matter the EFTofLSS is not trying to replace $N$-body simulations tout court. It is rather stressing a complementarity that might lead to sensible progress in the field. Since the EFTofLSS can provide analytic  control in the quasi-linear regime, simulations can focus on understanding the physics {\it within} the non-linear regime, where the EFTofLSS will have nothing to say. By focusing, at least quasi-entirely, on the short distance physics, simulation codes can afford higher resolution and therefore more accurately reproduce the correct physics.\\

The situation is worse when we deal with baryon physics. In this case, we currently do not have first-principles, at least-in-principle correct, simulation codes, but only codes that accurately implement models. Due to the huge range of scales necessary to simulate star formations in a cosmological setting, it is hard to imagine that we will have at our disposal, within a short time, a first-principles code for the simulation, say, of a galaxy. Since baryons physics affects long distances, it is sometimes necessary to simulate such effects on large boxes, at the cost of accuracy. The purpose of this paper is to provide an Effective Field Theory treatment for the baryonic effects in LSS in the quasi-linear regime, to complement simulations within the non-linear regime. 

The idea of the approach is very simple. It follows directly from the construction we did for dark matter, which we now briefly recap. In the EFTofLSS, since we do not focus on scales shorter than the non-linear scale, dark matter is described as a fluid-like system with an effective stress tensor. At a given order in perturbation theory, the stress tensor  is effectively described at large distances by a few parameters such as pressure, viscosity, etc.. We should explicitly mention that what makes the universe filled with dark matter describable in these terms is the fact that the relative displacement between two nearby dark matter particles is very small in the current universe, indeed of order the non-linear scale. 

Now, let us think about how baryons are different from dark matter. From the time of recombination to the  formation of the first stars, baryons and dark matter behave as the same species, with the only difference being their different initial conditions. Since the equations of motion are independent of the initial conditions, this means that the equations of motion are the same~\footnote{Truly, even in this situation there is an effect, that we check in this paper to be extremely small in our universe, due to the fact that different initial conditions for the same system leads to different non-linear structures, and therefore to different effective stress tensors.}. In particular, until the formation of the first stars, the EFT that describes baryons is identical, even in the parameters, to the one that describes dark matter. 

The story changes when stars begin to be formed. Supernovae explode, active galactic nuclei form, etc.. Very energetic radiation is emitted from the first stars, re-ionizing the baryons, which become hotter and must develop some form of pressure. Clearly, the story is very complicated.  The observation that we find crucial in order to develop our effective field theory is that, notwithstanding the vicissitudes of the star-formation physics, baryons do not move very much. The transfer of mass and momentum on scales longer than about the non-linear scale is negligible, as baryons are very non-relativistic, even when hot. This can indeed be checked by observing that, apart for order one numbers, in a cluster baryons and dark matter occupy the same regions. Therefore, the {\it same} EFT that describes dark matter can describe baryons as well, with the only difference that now baryons and dark matter are allowed to exchange momentum through gravity. Furthermore, the numerical coefficients that describe the size of the induced stress tensor are expected to be only order one different. This observation implies that the {\it functional form} of the effect of baryonic physics at long distances is fixed, and is independent of the details of the baryonic physics, and actually the same as the one we have for dark matter, apart for numerical prefactors. For example, in the power spectrum of baryons and of dark matter, the leading corrections from baryonic physics go as
\be\label{eq:intro}
\Delta P_b(k)\propto \left(\frac{k}{\knl}\right)^2 P_{11}^A(k)\ , \qquad \Delta P_c(k)\propto \left(\frac{k}{\knl}\right)^2 P_{11}^A(k) \ .
\ee
where $P_{11}^A$ is the linear power spectrum of the adiabatic mode $\delta_A=w_c \delta_c+w_b\delta_b$, with $\delta_{c,b}$ being the dark matter and baryon overdensities, $w_{b,c}$ being their relative contribution to the energy density of the universe, and $\knl$ is the wavenumber associated to the non-linear scale. Eq.~(\ref{eq:intro}) tells us that different models of star formation will lead at long distances to the same functional form for their corrections, apart for their overall size. This is interesting for two reasons. First, it tells us that we can in principle afford not to have a derivation of the size of these terms from first principles: since the functional form of the correction is known, we can fit it directly to observations. Of course, it is much better not to have to fit for any new parameter. So, \eqn{eq:intro} tells us that in order to reproduce the leading long distance information, simulation codes can simply work towards determining the numerical coefficients in \eqn{eq:intro}, which is probably an easier job than determining the full functional form. As we will verify, different star formation models will differ in the numerical value of the coefficients in (\ref{eq:intro}), but not in the functional form. Additionally, thanks to the EFTofLSS, simulation codes can begin to be run on smaller boxes, so that their accuracy can be increased. \\

After constructing the relevant EFT equations, in order to compare the solutions to simulation data, we need to take care of the effect of infrared modes.  In the current universe, it has long been realized that large infrared displacements harm the perturbative expansion and need to be resummed. In the case where there is only one fluid, general theorems~\cite{Scoccimarro:1995if,Carrasco:2013sva} tell us that  that these effects cancel for equal time power spectra, and therefore it is only the displacements from modes of order the BAO scale that need to be resummed~\cite{Senatore:2014via}. The reason why displacements induced by arbitrarily long modes do not contribute at least to some observables is because the displacement induced by long modes is proportional to the gradient of the Newtonian potential, which, by general relativity, is just a gauge artifact and does not affect local observables such as the equal time power spectra of short modes. However, in the case of two fluids, there is a relative displacement that cannot be set to zero by a gauge transformation, and that therefore gives rise to dynamical effect in all observables. This effect, first pointed out in~\cite{Tseliakhovich:2010bj}, is large in our universe at redshift of order $z\sim 40$ and leads to a breaking of perturbation theory~\cite{Bernardeau:2012aq}. Since this is an infrared effect, we generalize the formulas of~\cite{Senatore:2014via} to the case of two fluids, to provide a way to systematically resum such an effect in an analytic way.\\

Endowed with all these expressions, we are ready to compare with simulation data. We use two kinds of simulations. In the first ones, that we discuss in App.~\ref{app:darkmattersim}, baryons are simulated with all baryon effects shut off, but still keeping the different initial conditions that baryons and dark matter have in our universe. This gives us a measure of how much the different initial conditions are important, an effect that we confirm to be small at redshift zero.  Then, in Sec.~\ref{data}, we compare with simulations of baryonic physics. The EFTofLSS predicts that the effect should be described by the functional form of (\ref{eq:intro}), up to a scale which can also be estimated, given by when higher order corrections become relevant. As we will see, the comparison seems to work extremely well.

%
%
%
%

\section{Equations of Motion and Perturbative Solutions}\label{eom}

\subsection{Equations of Motion}

In the Eulerian description, the EFTofLSS for dark matter takes the form of fluid-like equations.  The generalization of the equation of motions to two species is straightforward. The equations are not exactly the ones of two fluids because the EFT is non-local in time~\cite{Carrasco:2013mua,Carroll:2013oxa}. This is because all modes of interest, including the UV modes that have been integrated out, evolve on Hubble time scales. Baryons and dark matter conserve their number density, but exchange momentum through gravity. This means that the effect of short distance physics on dark matter and baryons does not appear in the form of an effective stress tensor. More in detail,  we write 
\bea
&&\nabla^2\phi = \fr{3}{2} H_0^2 \fr{a_0^3}{a} (\Omega_c \delta_c+\Omega_b \delta_b) \n
&&\dot \delta_{c} = -\fr{1}{a}\partial_i ((1+\delta_{c})v_{c}^i) \n
&&\dot \delta_{b} = -\fr{1}{a}\partial_i ((1+\delta_{b})v_{b}^i) \n
&&\partial_i \dot v^i_{c}+H \partial_i v^i_{c}+\fr{1}{a} \partial_i(v^j_{c} \partial_j v^i_{c})+\fr{1}{a}\partial^2 \phi = - \fr{1}{a} \partial_i \left( \partial \tau_\rho \right)_{c}^i+  \fr{1}{a} \partial_i (\gamma)_c^i \ , \n
&&\partial_i \dot v^i_{b}+H \partial_i v^i_{b}+\fr{1}{a} \partial_i(v^j_{b} \partial_j v^i_{b})+\fr{1}{a}\partial^2 \phi = - \fr{1}{a} \partial_i \left( \partial \tau_\rho \right)_{b}^i+  \fr{1}{a} \partial_i (\gamma)_b^i \ ,
\label{fluid}
\eea
where 
\be \label{stresstensor}
\left( \partial \tau_\rho \right) _{\sigma}^i = \frac{1}{\rho_\sigma} \partial_j   \tau^{ij}_\sigma \ ,\qquad (\gamma)_c^i =\frac{1}{\rho_c} V^i \ ,\qquad (\gamma)_b^i =-\frac{1}{\rho_b} V^i \ .
\ee
Here $\sigma=c,b$, and all the fields that appear in this paper are the long wavelength fields defined in~\cite{Carrasco:2012cv} unless otherwise stated.  
The $\Omega_\sigma$ parameters are the present day energy fractions of the various components of the universe, and we will frequently make use of the following definitions:
\bea
&& \Omega_m = \Omega_c + \Omega_b \n
&& w_b=\fr{\Omega_b}{\Omega_m} \  \ , \  \  w_c=\fr{\Omega_c}{\Omega_m}  \ .
\eea
$\tau^{ij}_\sigma$ is the effective stress tensor that comes from integrating out the short distance physics. However, while in the single species case the contribution from short distance physics can be entirely encapsulated in an effective stress tensor, this is not so in the case of two gravitationally interacting fluids, where a new term is necessary, which we call $V^i$. In fact, if the effect of short distance physics were to be described solely by an effective stress tensor, the momentum of each of the two species would be conserved. But this is clearly not the case as gravitational interactions exchange momentum between the two species. This means that in the momentum equation the effect of short distance fluctuations requires the addition of a new term: $V^i$. Notice that since interactions conserve the identity of the particles, the continuity equations for both species do not require any new term. Furthermore, the total momentum, sum of the two momenta of the species, is conserved, and therefore the two new terms $(\gamma^i)_c$ and $(\gamma^i)_b$ are such that they cancel when considering the total momentum~\footnote{Following~\cite{Baumann:2010tm}, it is possible to explicitly identify the short distance term that cannot be encapsulated in a stress tensor. Since dark matter is collisionless, and baryon-baryon interactions conserve momentum, the new term must arise from the gravitational ones. Let us start with a single fluid. We can focus on the short distance term $\rho_s \pd_i \phi_s$ and show how it can be rearranged into an effective energy tensor term $\pd_j\tau^{ij}$: 
\bea
\rho_s \pd_i\phi_s=\frac{\pd^2\phi_s}{4\pi G} \pd_i\phi_s=\frac{1}{4\pi G} \pd_j\left( \pd_i\phi_s \pd_j\phi_s-\frac{1}{2} \delta_{ij} (\pd_l\phi_s\pd^l\phi_s)\right)\ ,
\eea 
where in the second passage we have used the Poisson equation for a single species: $\pd^2\phi_s=4\pi G \rho_s$. In the case of two species, the Poisson equation changes to $\pd^2\phi_s=4\pi G (\rho_{c,s}+\rho_{b,s})$, so that in the momentum equation for dark matter we have
\bea
\rho_{c,s} \pd_i\phi_s=\left[\frac{\pd^2\phi_s}{4\pi G}-\rho_{b,s}\right] \pd_i\phi_s=\frac{1}{4\pi G}\left[ \pd_j\left( \pd_i\phi_s \pd_j\phi_s-\frac{1}{2} \delta_{ij} (\pd_l\phi_s\pd^l\phi_s)\right)-\rho_{b,s}\pd_i\phi_s\right]\ .
\eea 
We see that the fact that the two species couple to the same gravitational potential, which is nothing but the equivalence principle,  leads to a term, proportional to $\rho_{b,s}\pd\phi_s$, that does not take the form of a stress tensor, which is a total derivative.  This term is exactly the term, with opposite sign, that appears directly in the equation for the baryon momentum, so that it cancels in the equation for the total momentum. We also see that this term is proportional to $\rho_b$: in the limit of no baryons, we are back to a single fluid and just to an effective stress tensor.
}.

Because the EFTofLSS is non-local in time, the response of the terms in Eq.~(\ref{stresstensor}) to the long wavelength fields will be an integral over some kernel in time of an expansion in powers and derivatives of $\partial_i \partial_j \phi$, $\partial_i v_\sigma^j$, evaluated along the flow.  The lowest order term in this expansion we have
\bea
&&-\left( \partial \tau_{\rho} \right)^i_\sigma ( a , \bx) +(\gamma)^i_\sigma ( a , \bx) =  \\ \nonumber
&&\qquad\qquad\int d a' \left[\kappa_\sigma^{(1)}( a , a' )\,    \partial^i \partial^2 \phi ( a' , \bx_{\rm fl}( \bx ; a,a') ) +\kappa_\sigma^{(2)}( a , a' )\, \fr{1}{H}\partial^i \partial_j v^j_\sigma ( a' , \bx_{\rm fl}( \bx ; a,a') )   \ldots \right ] \ ,
\label{str}
\eea
where $\kappa$ is the kernel and the fluid line element $\bx_{\rm fl}$ is defined implicitly as~\cite{Carrasco:2013mua} 
\be
\bx_{\rm fl} ( \bx ; a , a' ) = \bx - \int_{a'}^a d a''  \;\frac{d \tau }{d a} ( a'' ) \; \bv ( a'' , \bx_{\rm fl} ( \bx ; a , a'') )\ ,
\ee 
where $\tau$ is conformal time.
These terms  associated to the past trajectory appear at high order in the fluctuations. In this paper we will focus on calculations done at one loop level, where it is sufficient to evaluate these counterterms on the linear solutions. In this way we can use the fact that several terms have the same functional form at low orders in perturbation theory. 
At this order, the non-locality in time corresponds simply to a redefinition of the parameters of a would-be local-in-time theory~\cite{Carrasco:2013mua}. In fact, using the linear solutions to \eqn{fluid}, we can schematically write
\be
-  \left( \partial \tau_\rho \right )^i ( a , \bx)  +(\gamma)^i_\sigma ( a , \bx) \sim  \left( \int d a' K (a , a' ) \frac{D(a')}{D(a_0)}  \right)  \partial_i \delta( a , \bx)  \ ,
 \ee
where, at linear order in the perturbations, we have neglected a factor of $e^{i \bk \cdot \left( \bx_{\rm fl} - \bx \right)}\simeq 1$.
We can symbolically perform the integral over $a'$ and are left with just a function of one variable $a$, which we use to define the local-in-time speed-of-sound-like parameters as follows 
\bea  \label{eq:effectivestress}
&&\partial_i \left( \partial \tau_\rho \right)^i_{c}-\pd_i(\gamma)^i_c ( a , \bx)=\\ \nonumber
&&\qquad\qquad    (2\pi)\,9\,c_{c,g}^2(a) \frac{H^2}{k^2_{NL}}  \left ( w_c  \partial^2\delta_c +w_b  \partial^2\delta_b \right )+ (2\pi)  \,9\,c_{c,v}^2(a) \frac{ H^2}{k^2_{NL}} \, \partial^2  \delta_c+ \ldots\ ,  \\ \nonumber
&&\partial_i \left( \partial \tau_\rho \right)^i_{b} -\pd_i(\gamma)^i_b ( a , \bx)= \\ \nonumber
&&\qquad\qquad   (2\pi) \,9\,c_{b,g}^2(a)\frac{ H^2}{k^2_{NL}}  \left ( w_c  \partial^2\delta_c +w_b  \partial^2\delta_b \right )+  (2\pi)\,9\,( c_{b,v}^2(a)+c_\star^2(a))\frac{ H^2}{k^2_{NL}} \, \partial^2  \delta_b  + \ldots  ,
\label{stress}
\eea 
where the ellipsis represents terms that are either higher order in $\delta_{c,b},\ldots$, or higher derivatives of $\delta_{c,b},\ldots$, or stochastic terms, all of which are negligible at the order we work in this paper.  

Let us explain in some detail the structure of the effective stress tensor above in (\ref{eq:effectivestress}), where we have included only the leading terms. The terms in $c_{c,g}^2$ and $c_{b,g}^2$ can be intuitively called the gravitationally induced (unitless) speed of sound parameters for the dark matter and baryons respectively because they come from the $\pd^2\phi$ term in \eqn{str}. This fixes the dependence on the relative abundances $w_c$ and $w_b$. The terms $c_{c,v}^2$ and $c_{b,v}^2$ are the response of the stress tensor of dark matter and baryons to the respective gradients of the velocity fields $\pd_i v^i_\sigma$, after substituting for the continuity equations. Finally, the term in $c_\star^2$ is a speed of sound that is induced by baryonic, or star formation, physics (as the subscript $_\star$ clearly indicates). The differences between $c_{c,g}^2$ and $c_{b,g}^2$, as well as between $c_{c,v}^2$ and $c_{b,v}^2$, are just due to baryonic physics, and therefore are expected to be of the same order as  $c_\star^2$. According to how star formation proceeds, this can be a number much smaller, or much larger than one, and we will later verify with comparison with simulations that its size seems to be  somewhat smaller than one. Following the convention of~\cite{Carrasco:2013mua} for the linear power spectrum and the definitions of the coupling constant, the factor of $(2\pi)\,9 $ have been chosen so that all the remaining numbers are expected to be order one.

The structure of the effective stress tensor is heavily affected by the fact that we are dealing with two fluid-like species that interact only gravitationally among each other. The fact that the interaction is only gravitational means that the effect of one component on the other is mediated only by gravity. This enters in two ways: first, on what the curvature $\pd^2\phi$ is, but also on what the local inertial frame, determined by $\pd\phi$, is. If we go to the particular local inertial frame that is the center of mass frame, we will have that, contrary to the single-species case, the velocity of each species is non vanishing. In this frame there is a relative velocity surviving for the two species
\bea
&&v_{c,{\rm CM}}= v_c-\left(w_c v_c+w_b v_b\right)=w_b (v_c- v_b)\ ,\\ \nonumber
&&v_{b,{\rm CM}}=v_b-\left(w_c v_c+w_b v_b\right)=w_c (v_b- v_c)\ .
\eea
Notice that in the limit $w_b\ll1$, $v_{c,{\rm CM}}\to 0$, $v_{b,{\rm CM}}\to v_b-v_c$, as it is quite intuitive. At least in principle, the effective stress tensor can now depend directly on these fields. Notice that if we were to allow for such a term to appear in the stress tensor, without any additional suppression, it could lead to an order one effect on the linear equations.  However, units and indices (and even physical intuition), come to our rescue. In fact, if $L$ represents units of length and $T$ units of time, $ \left( \partial \tau_\rho \right)^i$ has units of $1/L$. The only combination linear in $v$ that we can write  with these units is $\pd^2 v^i/H$,  but this term has two derivatives acting on $v$, and indeed we have already included it. If we want a derivative not to act on the velocity, then we need to go to quadratic terms such as $ v^i_{\rm \sigma, CM} \pd^2\delta/H$. We conclude that at quadratic level velocities without a derivative acting on them do not appear only through the dependence of $x_{\rm fl}$ on $v$, but also in these other combinations.  Since these terms are subleading at the order we work at, we neglect them for the current paper.

By substituting \eqn{eq:effectivestress} in \eqn{fluid}, we can find the effective Eulerian equations of motion. After Fourier transforming and changing time variables from $t$ to $a$, the fluid-like equations become
\bea \label{equation}
&&a \cH \delta_c'   +\theta_c = -\alpha_{cc} \\ \nonumber
&&a \cH \delta_b' +\theta_b = - \alpha_{bb} \\ \nonumber
&&a \cH \theta_c' +\cH \theta_c+\fr{3 \cH_0^2 a_0^3}{2a} (\Omega_c \delta_c+\Omega_b \delta_b) = -\beta_{cc} +   (2\pi)\,9\, c_{c,g}^2(a) H^2 \frac{k^2}{k^2_{NL}} \left ( w_c  \delta_c +w_b  \delta_b \right )\\ \nonumber
&&\qquad\qquad\qquad \qquad\qquad \qquad\qquad\qquad+  (2\pi)\,9\, c_{c,v}^2(a) H^2  \frac{k^2 }{k^2_{NL}} \delta_c  \ ,\\ \nonumber
 &&a \cH \theta_b'+\cH \theta_b+\fr{3 \cH_0^2 a_0^3}{2a} (\Omega_c \delta_c+\Omega_b \delta_b) = -\beta_{bb} +   (2\pi) \,9\,c_{b,g}^2(a) H^2 \frac{k^2 }{k^2_{NL}}  \left ( w_c  \delta_c +w_b  \delta_b \right )  \\ \nonumber
&&\qquad \qquad\qquad \qquad\qquad \qquad\qquad\qquad+  (2\pi) \,9\,\left(c_{b,v}^2(a)+c_\star^2(a)\right) H^2 \frac{k^2 }{k^2_{NL}}  \delta_b \ ,
\eea  
where 
\bea
&&\alpha_{\sigma \kappa} \equiv \int \fr{d^3 q}{(2\pi)^3} \fr{\kv \cdot \qv}{q^2}  \ \delta_{\sigma}(\kv-\qv)\theta_{\kappa}(\qv) \n
&&\beta_{\sigma \kappa} \equiv \int \fr{d^3 q}{(2\pi)^3} \frac{ k^2 \qv \cdot (\kv-\qv)}{2 q^2(\kv-\qv)^2} \  \theta_{\sigma}(\kv-\qv)\theta_{\kappa}(\qv) \ .
\eea

It will often be convenient to transform to the basis of adiabatic and isocurvature modes, which are defined by 
\bea
\label{basis}
\delta_A &\equiv& w_c \delta_c + w_b \delta_b \n
\delta_I &\equiv& \delta_c - \delta_b \ ,
\eea
and the same for the $\theta$ variables. The adiabatic mode is the total density fluctuation, and the isocurvature mode is the relative density fluctuation.

It is worth to explicitly discuss the expansion parameters that control the perturbative expansion in the case of two species. There are five parameters which are
\bea\label{eq:epsparameters}
&&\epsilon_{\delta <} = \int^k_0 {d^3k' \over (2 \pi)^3} P_{11}(k')\ , \\ \nonumber
&&\epsilon_{s >} =k^2  \int_k^\infty {d^3k' \over (2 \pi)^3}  {P_{11}(k') \over k'^2}\ , \\ \nonumber
&&\epsilon_{s_<} =k^2  \int_0^k {d^3k' \over (2 \pi)^3}  {P_{11}(k') \over k'^2}\ ,
\eea
and
\bea\label{eq:epsrelparameters}
\epsilon_{s < }^{\rm rel}(k) = k^2  \int_0^k \frac{ d^3 k' }{(2\pi)^3} \frac{\tilde P_{11}(k')}{k'{}^2} \ ,\\ \nonumber
\epsilon_{s > }^{\rm rel}(k) =  k^2 \int_k^\infty \frac{ d^3 k' }{(2\pi)^3} \frac{\tilde P_{11}(k')}{k'{}^2} \ .
\eea
The first three parameters are the same that appear in the dark matter only case~\cite{Porto:2013qua}. The first represents the effect on a given mode $k$ of tidal forces from longer modes. The second represent the effect on the same mode $k$ of displacements induced by shorter modes. The third represents the effect on the same mode $k$ of displacements from longer modes. The next two parameter appear because we have two species. The first   represents the effect on the same mode $k$ of the relative displacements induced by longer modes. $\tilde P_{11}(k)$ is indeed the power spectrum of the log-derivative with respect to the scale factor $a$ of difference in the dark matter and baryon overdensities: $\pd(\delta_c-\delta_b)/\pd\log a$.  The last parameter is the effect on a mode $k$ of the relative displacements induced by short modes.

An Eulerian calculation in the EFTofLSS amounts to perturbatively expanding in all of these parameters. A Lagrangian calculation does not expand in $\epsilon_{s < }$ and $\epsilon_{s < }^{\rm rel}$, and it is therefore a better approach. Unfortunately, calculations done using the Lagrangian approach can be tedious~\footnote{At least to some of us.}, and more subtle when the renormalization procedure is not straightforward~\cite{Porto:2013qua}. However, it is possible to obtain the non-perturbative result in $\epsilon_{s<}$ by performing suitable manipulation of the Eulerian calculations. This was shown to be possible in~\cite{Senatore:2014via}, and we will generalize it here to the case of two species, which allows us to treat non-perturbatively $\epsilon_{s<}$ and $\epsilon_{s < }^{\rm rel}$ as well.

%
%
%

\subsection{Linear Solution} \label{linear}

First we need to find the linear solution to the two-fluid equations of motion \eqn{equation}. This is similar to the one-fluid case because the linear equations (including setting to zero the counterterms) are diagonal in the adiabatic-isocurvature basis of \eqn{basis}.

In this basis the linear equations are
\bea
&& - a^2 \cH^2 \dfpp_A  - \left( 2 a \cH^2 + a^2 \cH \cH' \right) \dfp_A + \fr{3 \cH_0^2 a_0^3 \Omega_m }{2a} \df_A = 0 \n
&& - a^2 \cH^2 \dfpp_I  - \left( 2 a \cH^2 + a^2 \cH \cH' \right) \dfp_I = 0 \ .
\label{isolin}
\eea
There are two solutions to each of these equations, but the solutions that grow the fastest with $a$ quickly dominate. The dominant solution to the equation for the isocurvature mode is constant in time, so $\df_I ( k , a ) = \delta_I(k)$. The dominant solution of the adiabatic equation is instead $\df_A( k , a) = D(a) \delta_A (k) / D(a_0) $, where $D(a) \propto \left( \cH (a) / a \right) \int_0^a da' \cH(a')^{-3} $ is called the linear growth factor and is the fastest growing solution to
\be \label{apple}
- a^2 \cH^2 D''(a) - \left( 2 a \cH^2 + a^2 \cH \cH' \right) D'(a) + \fr{3 \cH_0^2 a_0^3 \Omega_m }{2a} D(a) = 0 \ .  
\ee
Although most of the time we will be interested in ratios of growth factors $D$'s, the conventional normalization is $D(a_0)=1$.  For the initial conditions of the linear solutions we use the present-day linear power spectrum which can be taken from CAMB~\cite{Lewis:1999bs}.

Because of the different evolution of the adiabatic and isocurvature modes, the current ratio of $\df_I / \df_A$ scales as
\be
\frac{\df_I ( k , a_0 ) }{\df_A ( k , a_0)} \sim \left ( \frac{ D(a_i) }{ D(a_0)} \right ) \frac{\df_I ( k , a_i ) }{ \df_A ( k , a_i)}\ ,
\ee
where $a_i$ is some early time. The isocurvature modes become more suppressed with time, and, as a result, the current ratio is about $\df_I ( k , a_0 ) / \df_A ( k , a_0)\sim 10^{-2}$. Since we aim at doing calculations at percent level accuracy, this tells us that we must keep the isocurvature mode at tree level, i.e. for the linear solution, but can neglect it inside loops.

%
%
%
%

\subsection{One-loop Solution}
We now proceed to the solution to the EFT equations at one loop, for which we use \eqn{equation} with the counter-terms set to zero for now. In the adiabatic-isocurvature basis the fluid equations without counter-terms are:
\bea
&&a \cH \delta_A'   +\theta_A=-\alpha_{AA} -  \alpha_{II} w_b w_c  \n
&& a \cH \delta_I' +\theta_I =- \alpha_{AI}-\alpha_{IA}-\alpha_{II} (w_b-w_c) \n
&&a \cH \theta_A'  +\cH \theta_A+\fr{3 \cH_0^2}{2a} (\Omega_c +\Omega_b) \delta_A = -\beta_{AA}-\beta_{II} w_b w_c\n
&& a \cH \theta_I' +\cH \theta_I = -2\beta_{AI}-\beta_{II} (w_b-w_c)   \ .
\eea 
Notice that isocurvature modes at higher order in $\delta$ are always sourced by at least one lower order isocurvature mode, while adiabatic modes can be sourced by adiabatic modes alone. This means that any loops producing isocurvature modes are suppressed with respect to the corresponding loops producing adiabatic modes by at least $10^{-2}$, so they are subleading. Since $\delta_I/\delta_A \sim 10^{-2}$ at linear order, we only need to keep the linear isocurvature modes, and for one-loop calculations we will only include $\delta_A$.

Since the equations for the adiabatic mode neglecting isocurvature are exactly those for a single fluid with density $\Omega_m=\Omega_c +\Omega_b$, we can use the same method as \cite{Carrasco:2012cv} to solve them. We will use the EdS approximation so that the $a$ and $k$ dependence separates, which is exact for an $\Omega_m=1$ cosmology, and is correct to percent level in $\delta$ for $\Lambda$CDM cosmology (see for example \cite{Carrasco:2012cv}). To implement this approximation, we make the ansatz
\bea
&& \delta_b^{(n)}( k , a)  = \delta_c^{(n)}( k , a)  = D(a)^n A^{(n)} ( k ) \n
&& \theta_b^{(n)} ( k , a) =  \theta_c^{(n)} ( k , a) = - \cH f D(a)^n B^{(n)} (k) \ ,
\eea
where $f^2 = \fr{\cH_0^2}{ a\cH^2} (\Omega_c+\Omega_b)$.  The EdS approximation relies on $ ( \Omega_m \cH_0^2  a_0^3 / ( a \cH^2) ) / (a D' / D)^2 $ being close to unity.  This ratio is one at early times and is $1.15$ at $a=1$ \cite{Carrasco:2012cv}, but is close to one for most of the time evolution.  The fact that it is close to one for most of the time evolution allows the approximation to be accurate to percent level since gravitational clustering is not affected too much by the latest times.  

 Using this approximation, we can solve for $A^{(n)} ( k )$ and $B^{(n)}(k)$ algebraically.  In particular, this leads to  
\bea
&& 2  \la \delta_\sigma^{(1)} (\bk , a_0) \delta_\kappa^{(3)}( \bq , a_0) \ra_\Lambda = (2 \pi)^3 \delta( \bk + \bq ) P^A_{13} ( k , \Lambda)  \n
&& \la \delta_\sigma^{(2)} (\bk , a_0) \delta_\kappa^{(2)}( \bq , a_0) \ra_\Lambda = (2 \pi)^3 \delta( \bk + \bq ) P^A_{22} ( k , \Lambda) \ ,
\eea
where $\sigma = b , c$ and $P^A_{13}$ and $P^A_{22}$ are the standard single fluid kernels \footnote{Explicitly, the loop integrals are
 \bea
P^A_{22}(k, \Lambda)&&=\fr{k^3}{392 \pi^2}\int_0^{\Lambda/k}  dr \int_{-1}^1dx \fr{(-10 r x^2+3 r+7 x)^2}{(r^2-2 r x+1)^2}P^A_{11}(k r)P^A_{11}(k \sqrt{r^2-2 r x +1}) \n
P^A_{13}(k, \Lambda)&&=\fr{k^3}{1008 \pi^2}P^A_{11}( k ) \n
&&\int_0^{\Lambda/k} dr \left( \fr{3}{r^3}(r^2 - 1)^3(7 r^2+2)  {\rm log } \Bigl  | \fr{1+r}{1-r} \Bigr  | -42 r^4+100r^2+\fr{12}{r^2}-158 \right) P^A_{11}(kr) \ .
\label{loops}
\eea
Notice that the power spectra generally need to be smoothed over (or cut off at) a scale $\Lambda$, because the UV theory at energies greater than $\Lambda$ is not under perturbative control.  However, the final result will be $\Lambda$-independent because the $\Lambda$-dependence of the $c^2$ counter-term parameters precisely cancel the leading $\Lambda$-dependence of the loop integrals by construction. In principle there is residual $\Lambda$ dependence of powers of $k/\Lambda$ in the loops due to the higher-derivative corrections we have neglected in the effective stress tensor. However, these effects go to zero in the decoupling limit $\Lambda \rightarrow \infty$ after the terms that are divergent in $\Lambda$ have been renormalized.  From now on we will take the limit $\Lambda \rightarrow \infty$ and drop the $\Lambda$ dependence in the power spectra, which is consistent if, in the counter-terms we use, the $c^2$ parameters are calculated at $\Lambda \rightarrow \infty$.
}.

%
%
%
%

\subsection{Counter-terms}

As discussed in the EFTofLSS \cite{Carrasco:2012cv}, the coefficients of the counter-terms generally have two contributions: $c^2 ( a , \Lambda ) = c^2_{\rm finite} ( a , k_{\rm renorm} ) + c^2_{\rm ct}( a , \Lambda)$.  The $\Lambda$-dependent piece is responsible for canceling any divergences in one-loop diagrams as $\Lambda \rightarrow \infty$, so the diagram involving $c^2_{\rm ct}$ must have the same time dependence as the respective loop diagram.  There is no such constraint on the time-dependence of the finite part, though.  Furthermore, the loop diagrams considered in this paper are finite as $\Lambda \rightarrow \infty$, so the contribution of $c^2_{\rm ct}$ is finite and can be absorbed into $c^2_{\rm finite}$.  This means that perturbation theory does not determine the time dependence of $c^2_{\rm finite}$, which is the piece relevant to our calculation.  Thus, we should assume a general time dependence for the $c^2(a)$ parameters appearing in Eqs. (\ref{stress}) that could be measured in $N$-body simulations and used as an input for the EFT. In practice, however, the time dependence can be reabsorbed into a rescaling of the parameters if we restrict to one-loop order and consider only one redshift, as we do here~\footnote{A study of the EFTofLSS as a function of redshift is in progress~\cite{simon}.}.

The equation we need to solve to find the counter-term contribution is
 \be
 -a^2 \cH^2 \delta_\sigma^{(ct)}{}'' - ( 2 a  \cH^2 + a^2 \cH \cH' ) \delta_\sigma^{(ct)}{}' + \frac{ 3 \cH_0^2 a_0^3 \Omega_m}{2 a} \left( w_c \delta_c^{(ct)} + w_b \delta_b^{(ct)} \right) =  (2\pi)\,9\, c^2_\sigma ( a) \frac{\cH^2}{a^2} \frac{k^2}{\knl^2} \frac{D(a)}{D(a_0)} \delta_A^{(1)} ( k ) ,
 \label{cteq}
\ee
where we have redefined the counter-term coefficients in \eqn{sound} below, and left off the isocurvature mode on the right hand side because it is subleading.

We find that the solution to \eqn{cteq} results in the following contributions to the power spectra:
\bea
&& P^c_{\rm counter} ( k ) = -  2 (2\pi) \left( \bar{c}_A^2 ( a_0 ) + w_b \bar{c}_I^2 ( a_0 )  \right) k^2 P^A_{11} ( k ) \n
&& P^b_{\rm counter}( k ) = - 2  (2\pi)\left(  \bar{c}_A^2 ( a_0 ) - w_c \bar{c}_I^2 (a_0 ) \right) k^2 P^A_{11} ( k ) \n
&& P^{bc}_{\rm counter} = \frac{1}{2} \left( P^c_{\rm counter} + P^b_{\rm counter} \right) \ ,
\eea
where $2 \langle \delta^{(1)}_\sigma ( \bk , a_0 ) \delta^{(\rm ct)}_\sigma ( \bq , a_0 ) \rangle = ( 2 \pi )^3 \delta( \bk + \bq) P^\sigma_{\rm counter} ( k ) $ and for convenience we have defined new parameters $ \bar{c}_A$ and $ \bar{c}_I$\footnote{The counter-term parameters in \eqn{stress} only come into the equations of motion at one loop in the following two combinations:
 \bea
c^2_c ( a) =    c_{c,g}^2+c_{c,v}^2\ ,     \hspace{.5in}  c^2_b ( a) =   c_{b,g}^2+c_{b,v}^2 +c_\star^2   \ .
\label{sound}
\eea  
In the power spectra, the relevant parameters are the following integrals:
 \bea\label{eq:green}
&& \bar{c}_A^2 (a ) = \int^a d a' G_A ( a , a' ) \frac{D(a' ) }{D(a_0) }\,9\,   \frac{H^2}{\knl^2} \left( w_b c_b^2 (a') + w_c c_c^2 ( a' ) \right) \n
&& \bar{c}_I^2 ( a ) = \int^a d a' G_I (a , a' ) \frac{D(a' ) }{D(a_0) } \,9\,    \frac{H^2}{\knl^2}  \left( c_c^2( a' ) -  c_b^2 (a') \right) \ , 
\label{csai}
\eea 
where $ G_A $ and $G_I  $ are the retarded Green's functions for the linear equations 
\bea
&& - a^2 \cH^2 G_A''  - \left( 2 a \cH^2 + a^2 \cH \cH' \right)G_A' + \fr{3 \cH_0^2 a_0^3 \Omega_m }{2a}G_A = \delta^{(1)}(a-\tilde a) \ ,\n
&& - a^2 \cH^2 G_I''  - \left( 2 a \cH^2 + a^2 \cH \cH' \right) G_I' = \delta^{(1)}(a-\tilde a) \, \\ \nonumber
&& G_{A,I}(a,a)=0\ , \qquad \left.\pd_aG_{A,I}(a,\tilde a)\right|_{ a=\tilde a}=\frac{1}{\tilde a^2\cH(\tilde a)^2}\ .
\eea
Approximating the integrals in (\ref{eq:green}) with the corresponding expressions in EdS, and by choosing the time dependence of $c_s^2$ to be $\propto a^4$, just as an indication, we obtain
\bea\label{eq:simple_matching}
&&\bar{c}_A^2 (a _0) \simeq \left( w_b c_b^2 (a_0) + w_c c_c^2 ( a_0) \right)k_{\rm NL}^{-2}\ , \\ \nonumber
&& \bar{c}_I^2 ( a_0 ) \simeq \frac{18}{21}\left( c_c^2( a_0 ) -  c_b^2 (a_0) \right) k_{\rm NL}^{-2}\ .
\eea
This explains the factor of 9 that we included in the definition of $c_c^2$ and $c_b^2$ in (\ref{equation}).
}. We find that for the one-loop, equal-time power spectra of dark matter and baryons, the only inputs to the EFT are two time-independent parameters. Thus baryons are easily included in the EFTofLSS with only one additional parameter.

This is a good place to comment on the stochastic terms. We have two kind of stochastic terms. The stochastic term in $k^i(\gamma)^i_\sigma$ contributes to the power spectrum as $ \frac{w_b}{\knl^3}\left(\frac{k}{\knl}\right)^2$, as we expect that the correlation function of $(\gamma)^i_\sigma$ to be $k$-independent (Poisson-like). Even though suppressed by $w_b$, this term is less suppressed in terms of $k/\knl$ than the stochastic term of the effective stress tensor,   $k^ik^i\Delta\tau_{ij}$, which, similarly, contributes  as $\frac{1}{\knl^3}\left(\frac{k}{\knl}\right)^4$. The difference in power is associated to the violation of momentum conservation for a single species. Both of these terms are very subleading with respect to the contribution of the $c$-counterterms, which go as $\left(\frac{k}{\knl}\right)^2P_{11}(k)\sim \frac{1}{\knl^3}\left(\frac{k}{\knl}\right)^{0.5} $ for the current $k$'s of interest, and so we can safely neglect them.

%
%
%
%

\subsection{Summary of Expressions} \label{finaleq}
Let us recap for later convenience the results from this section. In this paper, we only concern ourselves with density-density power spectra, defined by $\langle \delta_\sigma ( \kvec , a_0 ) \delta_\sigma ( \kvec' , a_0 ) \rangle = (2 \pi)^3 \delta_D ( \kvec + \kvec' ) P^\sigma ( k )$. The power spectra for baryons and dark matter at one loop in the EFTofLSS are given by the following expressions
\bea \label{powerspectra}
P^{c}(k) &=& P^{c}_{11}(k) + P^A_{\rm 1-loop} (k)  -  2 (2\pi) \left( \bar{c}_A^2 ( a_0 ) + w_b \bar{c}_I^2 ( a_0 )  \right) k^2 P^A_{11} ( k )  \n
P^{b}(k) &=& P^{b}_{11}(k) + P^A_{\rm 1-loop} (k)- 2 (2\pi) \left(  \bar{c}_A^2 ( a_0 ) - w_c \bar{c}_I^2 (a_0 ) \right) k^2 P^A_{11} ( k )  \ ,
\eea
to one loop order, where $P^A_{\rm 1-loop} \equiv P^A_{22} + P^A_{13}$ and $ P^A_{22}$ and $P^A_{13}$ are given by Eq.(\ref{loops}).  We are also concerned with the \emph{total} matter power spectrum, which in our notation is called the adiabatic power spectrum, and is given by
\bea
P^{A } (k) &\equiv& w_c^2 P^c + 2 w_c w_b P^{bc} + w_b^2 P^b \n
&=& P^{A}_{11}(k) +  P^A_{\rm 1-loop}(k)  - 2 (2\pi)   \bar{c}_A^2 ( a_0 ) k^2 P^A_{11} ( k ) \ .
\eea
Notice that at linear level $\langle \delta^{(1)}_b ( \kvec , a_0 ) \delta^{(1)}_c ( \kvec' , a_0 ) \rangle = (2 \pi)^3 \delta_D ( \kvec + \kvec' ) P^{bc}_{11} ( k )$ where $P^{bc}_{11} = \sqrt{ P^b_{11} P^c_{11} } $ because the initial linear dark matter and baryon fields are both proportional to the primordial curvature perturbation.

%
%
%
%
%
%

\section{Details of IR Resummation} \label{irresumdetails}

The derivation that we have presented in Section \ref{eom} is called the Eulerian approach because the basic degrees of freedom are the values of the fields $\delta_\sigma ( \vec{x} , t )$ and $\theta_\sigma ( \vec{x} , t)$ as local functions of space and time which satisfy fluid-like equations.  The Eulerian approach does not handle bulk motions adequately because its basic degrees of freedom are localized, and a large bulk flow can have a big effect on the perturbative expansion even if very little is happening dynamically.  This is the IR effect that Eulerian perturbation theory inadequately Taylor expands on.  

However, the method given in \cite{Senatore:2014via} directly uses the Eulerian results, including counter-term expressions, as input for the IR resummation of the displacements, so the above discussion is still entirely relevant.  For the equal-time matter power spectra that we compute in Section \ref{finaleq}, the main effect of the IR resummation is to correctly describe the BAO oscillations.  Suppose that we are concerned with the power spectrum at some wavenumber $k$.  Roughly, one can measure the value of the power spectrum at wavenumber $k$ by observing in a box in real space of side length~$1/k$.  Intuition using the equivalence principle says that the overall motion of that box produced by modes with wavelength longer than $1/k$ cannot have any physical effect on equal time matter correlators.  Thus, one would conclude that no modes $k'$ with $k' \ll k $ contribute at the scale $k$, i.e. that modes IR with respect to $k$ have no contribution.  However, the BAO peak in real space produces an oscillation in $k$ space with characteristic scale $\Delta k_{\rm oscillations} \sim 2 \pi / (100 \, \rm Mpc)$, but its non-linear smoothing is controlled by $k_{\rm width}\gg \Delta k_{\rm oscillations}$.  Thus, if $\Delta k_{\rm oscillations} < k<k_{\rm width}$, the box of size $1/k$ would not be big enough to describe the oscillations.  The IR resummation correctly describes bulk flows associated to modes~$k$, with~$\Delta k_{\rm oscillations }  < k<k_{\rm width}$, and is thus able to match the BAO oscillations.

 To get a sense of how the expansion parameters in \eqn{eq:epsparameters} appear in perturbation theory~\cite{Senatore:2014via}, consider the one-loop SPT equal-time integrals in \eqn{loops}.  In the asymptotic limit $k' \ll k$ we find
\be
P^A_{22} (k) + P^A_{13}(k) \propto P^A_{11}(k)  \edl \ ,
\ee 
and for $k' \gg k$ we find
\be
P^A_{13}(k) \propto P^A_{11}(k) \esg \ . 
\ee
Thus, modes that are IR with respect to $k$ affect the power spectrum through the variance of the density fluctuations (i.e. $\edl$), and modes that are UV with respect to $k$ affect the power spectrum through displacements (i.e. $\esg$).  The parameter $\esl$, describing the effect of IR displacements on the mode $k$, does not appear in the matter equal-time correlation function because it cancels in the limit $k' \ll k$. It is indeed present in both the one-loop terms: $P_{13}(k)\supset-\frac{2}{3}P_{11}(k)\esl\,,\ P_{22}(k)\supset\frac{2}{3}P_{11}(k)\esl  $.  However, this parameter is important in unequal time correlators, and, as we described in the paragraph above, in the intermediate regime $\Delta k_{{\rm oscillations}} < ~k' < k$, so a controlled treatment of $\esl$ is necessary to understand unequal time correlators and BAO oscillations.  This is the purpose of the IR resummation: since in our universe the size of the effect in $\esl$ is of order one,  it treats the IR displacement effects non-perturbatively.  When we include baryons in large scale structure calculations, the generalizations of $\esl$ will play the important role in describing the effects of long wavelength relative velocity.  

In the single-species case, the resummation is accomplished by~\cite{Senatore:2014via,Angulo:2013qp} 
\be
P ( k ; t_1 , t_2) |_N = \sum_{j=0}^N \int dk' \hat{M}_{||_{N-j}} ( k , k' ; t_1 , t_2 ) P_{ j} ( k'; t_1 , t_2)  \ ,
\ee
where $P_{ j} ( k' ; t_1, t_2 ) $ is the Eulerian result at $j$-th order (i.e. the $j$-th order term in both $\esl$, $\esg$ and $\edl$ counted as equal), and $P ( k ; t_1 , t_2 ) |_N$ is the resummed result up to order $N$ (i.e. to $N$-th order in $\edl$ and $\esg$ counted as equal, and to all orders in $\esl$).  Because both $\edl$ and $\esg$ are treated perturbatively, we refer to both of them as $\edl$.  $\hat{M}_{||_{N-j}}  ( k , k' ; t_1,t_2 ) $ is a weighting factor given by 
\be \label{mexpression}
  \hat{M}_{||_{N-j}}  ( k , k' ; t_1,t_2 ) = \frac{1}{2\pi^2} \int d^3 q   \frac{ k' \, \sin ( k'q ) }{q}  e^{i \kvec \cdot \qvec} F_{||_{N-j}} ( \qvec , \kvec ; t_1 , t_2 ) \ ,
\ee
and 
\be
F_{||_{N-j}} ( \qvec , \kvec ; t_1 , t_2 )  = K_0 ( \kvec, \qvec ; t_1 , t_2) \cdot \left(K_0^{-1} ( \kvec , \qvec ; t_1 , t_2 ) \right) \Big| \Big|_{N-j} \ . 
\ee
The notation $\big| \big|_N$ means the expansion of $\edl$, $\esg$ and $\esl$, counted as equal, up to order $N$, while~$\big|_N$ means expand only $\edl$ and $\esg$, counted as equal, up to order $N$ and keep $\esl$ non-perturbative.  The $K_0$ factor depends only on the linear, Lagrangian solution.  See Appendix A of \cite{Senatore:2014via} for more details.  The expression for $K_0$ in the single fluid case at equal times is
\be
K_0 ( \kvec , \qvec; t , t)  = \exp \left\{ - \half \left(  X( q ; t )_1 k^2  + Y(q ; t )_1 (\kvec \cdot \hat{q} )^2   \right)  \right\} \ ,
\ee
where $\hat{q}$ is the unit vector in the direction of $\qvec$, and 
\begin{align} \label{xandy}
X(q;t)_1 & = \frac{1}{2 \pi^2 } \int_0^\infty dk \exp \left( \frac{-k^2}{\Lambda^2_{\rm Resum} } \right) P_{ 11} ( k ; t ) \left( \frac{2 }{3} - 2 \frac{ j_1(k q) }{kq} \right) \\
Y(q;t)_1 & = \frac{1}{2 \pi^2 } \int_0^\infty dk \exp \left(  \frac{-k^2}{\Lambda^2_{\rm Resum} } \right) P_{ 11} ( k ; t ) \left( -2 \, j_0 ( kq )  + 6 \frac{ j_1(k q) }{kq} \right)   \ ,
\end{align}
where $j_i (x)$ is the spherical Bessel function of kind $i$, and $\Lambda_{\rm Resum}$ is the IR scale up to which we wish to resum the IR modes.  Any dependence on $\Lambda_{\rm Resum}$ represents the effects of the displacements which have not been resummed.   After the resummation, we are left with a new parameter $\tilde{\epsilon}_{s <}  \ll 1 < \esl$ that controls the residual dependence on the long displacements, implying that the residual dependence on $\Lambda_{\rm Resum}$ becomes weaker with each loop order.

%
%
%

\section{Effective Treatment of Bulk Relative Displacements} \label{hiratasection}
In this section, we use the IR resummation of ~\cite{Senatore:2014via,Angulo:2013qp} to describe the effect of advection in the case of two fluids first pointed out in~\cite{Tseliakhovich:2010bj}.  First we will set up the formalism for applying the IR resummation technique to the two fluid system. The method that we will present offers the advantage that it is a perturbative scheme that can reach the desired precision by simply performing well-defined higher-order calculations. Here we will stop at the first non-trivial order, but the extension to higher orders is well defined and quite straightforward. In this section, as an example, we will focus in calculating the total power spectrum, which we can write as $ P^A = w_c^2 P^c + 2 w_b w_c P^{b c} + w_b^2 P^b$ and then apply the IR resummation to each of the power spectra on the right hand side separately. We apply the resummation in the baryon-CDM basis because these are the physical particles that we need to follow in the fluid flow.  The EFT is formulated in the baryon-CDM basis because it arises from integrating out the UV effects of physical particles. One can take different linear combinations of the baryon and cold dark matter modes to find the adiabatic and isocurvature modes, which are useful for loop calculations, but there is no sense in which there are ``adiabatic" or  ``isocurvature'' particles.  This means that the Lagrangian procedure of going to the frame of the particle and calculating displacements from original positions, which is used for the IR summation, only makes sense in the baryon-CDM basis.  Because the resummation is a nonperturbative operation, this is not equivalent to doing the resummation in the adiabatic-isocurvature basis and linearly changing to the baryon-CDM basis.  This is the first time that we see that the two bases are not equivalent.

The resummation of $P^b$ and $P^c$ is a trivial application of the results in  \cite{Senatore:2014via}, but with the replacements $X(q ; t)_1 \rightarrow X_{\sigma}(q;t)_1$ and $Y(q ; t)_1 \rightarrow Y_{\sigma}(q;t)_1$, where 
\begin{align} \label{xandy1}
X_{\sigma}(q;t)_1 & = \frac{1}{2 \pi^2 } \int_0^\infty dk \exp \left( \frac{-k^2}{\Lambda^2_{\rm Resum} } \right) P_{ 11}^{\sigma} ( k ; t ) \left( \frac{2 }{3} - 2 \frac{ j_1(k q) }{kq} \right) \\
Y_{\sigma}(q;t)_1 & = \frac{1}{2 \pi^2 } \int_0^\infty dk \exp \left(  \frac{-k^2}{\Lambda^2_{\rm Resum} } \right) P_{11}^{\sigma} ( k ; t ) \left( -2 \, j_0 ( kq )  + 6 \frac{ j_1(k q) }{kq} \right) \ ,
\end{align}
and again $\sigma = b, c$.  However, the corresponding functions for the cross-correlation have an important difference.  The expressions relevant in the resummation of $P^{bc}$ are 
\begin{align}
X_{bc}(q;t)_1 & = \frac{1}{2 \pi^2 } \int_0^\infty dk \exp \left(  \frac{-k^2}{\Lambda^2_{\rm Resum} } \right) \left( \frac{1 }{3}P_{ 11}^{b} ( k ; t ) + \frac{1}{3} P_{ 11}^{c} ( k ; t )   - 2 P_{11}^{bc} ( k ; t )  \frac{ j_1(k q) }{kq} \right) \\
Y_{bc}(q;t)_1 & = \frac{1}{2 \pi^2 } \int_0^\infty dk \exp \left(  \frac{-k^2}{\Lambda^2_{\rm Resum} } \right) P_{ 11}^{bc} ( k ; t ) \left( -2 \, j_0 ( kq )  + 6 \frac{ j_1(k q) }{kq} \right)  \ . 
\end{align}
The above formulae for $X$ and $Y$ above come directly from computing the displacement at linear level in the Lagrangian approach.  Now, in the expression for $X_{bc}(q;t)_1$, contrary to $X_{\sigma}(q;t)_1$, there is no cancellation at small $q$ because the power spectra inside of the parentheses are slightly different.  This leads to a higher value of $X_{bc}(q;t)_1$ at low $q$, which after the Fourier transform, leads to a lower value of $P^{bc}$ at high $k$.  This difference matters only in the resummation of $P^{bc}$ because in $P^b$ and $P^c$ the IR effects cancel out as in the single-fluid case, but $P^{bc}$ is affected by the relative velocity, and the long-wavelength behavior of the baryons and the dark matter does not cancel out.  The final expressions for the IR-resummed, equal time, $\delta$-$\delta$ power spectra are 
\be \label{resumeq}
P^\alpha ( k ; t ) |_N = \sum_{j=0}^N \int dk' \, \hat{M}^\alpha_{||_{N-j}} ( k , k' ; t, t ) \,  P^\alpha_{ j} ( k'; t) \ ,
\ee
where $\alpha = b , c , bc$.  The quantity $P^\alpha_{ j} ( k'; t) $ is the Eulerian result at $j$-th order, $P^\alpha ( k ; t ) |_N $ is the IR-resummed result up to $N$-th order, and $\hat{M}^\alpha_{||_{N-j}}$ is like in \eqn{mexpression}, but using $X_\alpha$ and $Y_\alpha$.

The nonlinear effect of advection comes from the large relative velocity between the dark matter and baryon fluids at the time of recombination, $z = 1020$.  This was expressed in~\cite{Tseliakhovich:2010bj} by computing
\bea\label{eq:integral}
\la v_{bc}^2 ( \vec{x} ) \ra = \int \frac{d k }{k} \Delta_\zeta^2 \left( \frac{ T_{\theta_b} ( k,a ) - T_{\theta_c} (k,a) }{k} \right)^2 \equiv \int \frac{ dk }{k} \Delta_{v,bc}^2 ( k ,a) \ ,  
\eea
where $T_{\theta}(k)$ is the transfer function for the velocity divergence, $\theta^{(1)}( \kvec , a ) = T_{\theta} ( k ,a ) \zeta ( \kvec )$, $\zeta(\kvec)$ is the primordial curvature perturbation satisfying $ \langle \zeta(\kvec) \zeta( \kvec' ) \rangle' = 2 \pi^2 \Delta_\zeta^2 / k^3$, and $\Delta_{\zeta}^2 = 2.42 \times 10^{-9}$ is the amplitude of the primordial curvature perturbation.  The result is that the relative motion is faster than the speed of sound of baryons at this time, with Mach number $\mathcal{M} = \sqrt{v^2_{bc}} / c_{s, b} \sim 5$, and this supersonic motion results in the baryons' advection from the potential wells created by dark matter.  This effect is seen at $z=40$ as a suppression of the total matter power spectrum near the baryon Jeans scale, $k_J \approx 200$ ${\rm Mpc}^{-1}$.  In this section only, we will express wave numbers in units of $\rm Mpc^{-1}$, without the typical factor of $ h = H_0  / 100 \times \rm{ Mpc / ( km / s ) } $, to make contact with \cite{Tseliakhovich:2010bj}. 

The fact that the integral in (\ref{eq:integral}) is large, even at $z=40$, signals that linear theory is no longer appropriate.  Ref.~\cite{Bernardeau:2012aq} pointed out that the one-loop contribution to the total matter power spectrum becomes large near $k_J$, signaling a nonlinearity. This is because $\eslrel(k)$, which shows up in the loop calculation, and is related to $\la v_{bc}^2 ( \vec{x} ) \ra$ by  
\be
\eslrel(k) = \left( \frac{k}{\cH} \right)^2 \int_0^{k} \frac{ dk' }{k'} \Delta_{v,bc}^2 ( k' ,a) \ , 
\ee
becomes large.  Fig. \ref{epsilonrel} presents a plot of this function, and shows that it becomes of order unity near $k \approx k_J$.  Indeed, this is exactly the type of parameter that our technique is designed to resum, because it is a bulk displacement coming from a velocity power spectrum.  The related parameter that could analogously be defined, and that describes the size of the center of mass motion induced by very infrared modes, does not appear in the loop calculation when computing equal time matter correlation functions, because it cancels as in Equation (7) of~\cite{Senatore:2014via}.  That cancellation can be explained by the equivalence principle, since the effects of IR modes on equal-time overall  matter power spectra, for $k\ll\Delta k_{\rm oscillations}$, can be removed with a coordinate transformation~\cite{Carrasco:2013sva}.  However, there is no such argument for the IR modes associated with \emph{relative} velocity. 

Similarly, a new parameter that we expect to appear in perturbation theory is the effect of relative short mode displacements, parameterized by $\esgrel$.  In Fig.~\ref{epsilonrel} we show that this parameter is perturbatively small for the modes of interest. 

\begin{figure}[htb!]
\begin{center}
\includegraphics[scale=1.1]{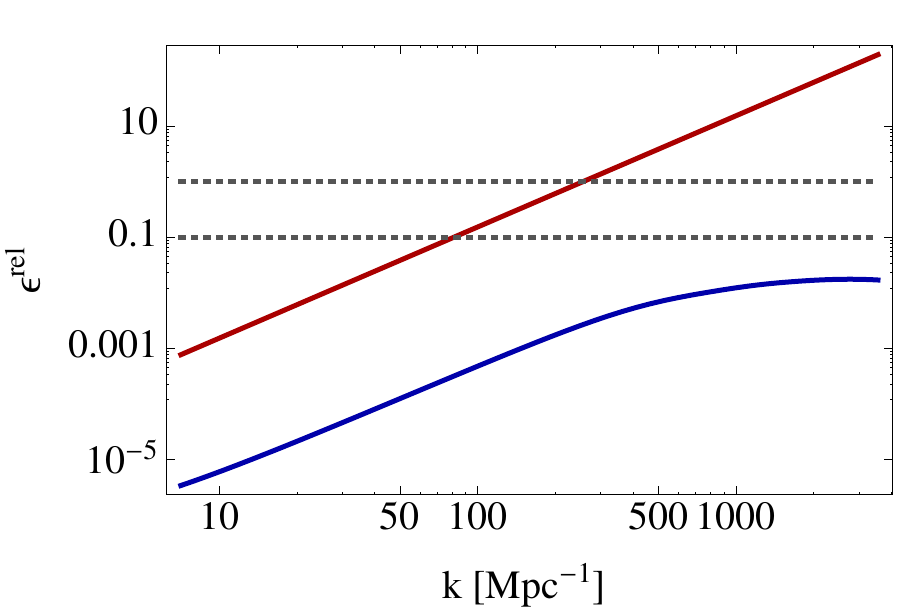}
\caption{ $\epsilon_{s<}^{\rm rel}$ and $\epsilon_{s>}^{\rm rel}$ at redshift $z=40$ are plotted in red and blue respectively. This plot shows that at redshift $z=40$ the effect of infrared relative motions becomes large at $k\approx k_J$, while the effect from short relative displacement is always perturbative.    }\label{epsilonrel}
\end{center}
\end{figure}

In Fig.~\ref{hirataplot} we show the result of our resummation for the quantity $\Delta_m^2  \equiv  k^3 P^A/(2 \pi^2) $ at $z=40$, where there is clearly a suppression of power near the baryon Jeans scale.  We only need to resum at tree level (the $j=0$ term in Eq. (\ref{resumeq})) since all of the $\epsilon_{\delta<}$ and $\epsilon_{s>}$ parameters are small.  Once we resum the relative velocity, the theory is extremely linear, as can be seen in Fig.~\ref{hirataplot}, where $\Delta_m^2 < 0.05$, and also from Fig.~\ref{epsilonrel}, where the new parameter $\epsilon_{s > }^{\rm rel}(k) $ is also shown to be small at the $k$'s of interest.  Fig.~\ref{hirataplot}  also shows that the effect comes almost entirely from the resummation of the cross correlation $\la \delta_b \delta_c \ra$, as expected from the fact that $\epsilon_{s<}^{\rm rel}$ becomes order unity.  We see that our calculation gives a similar suppression around $ k = 200  \, {\rm Mpc}^{-1}$ to the one in \cite{Tseliakhovich:2010bj} and \cite{Bernardeau:2012aq}, but it is qualitatively different because there are two bumps.  Also, our result starts to differ from the linear result starting at about $k = 20 \,  \rm Mpc^{-1}$, while the other studies show the difference starting around $k = 40 \,  \rm Mpc^{-1}$. The methods of~\cite{Tseliakhovich:2010bj} and~\cite{Bernardeau:2012aq} capture much of the effect of the relative velocity by effectively expanding perturbation theory around a bulk velocity, but our calculation does this in what we believe is a rigorous way, correctly taking into account gradient terms and organizing the result in a manifestly convergent perturbation theory.
\begin{figure}[htb!]
\begin{center}
\includegraphics[scale=1.2]{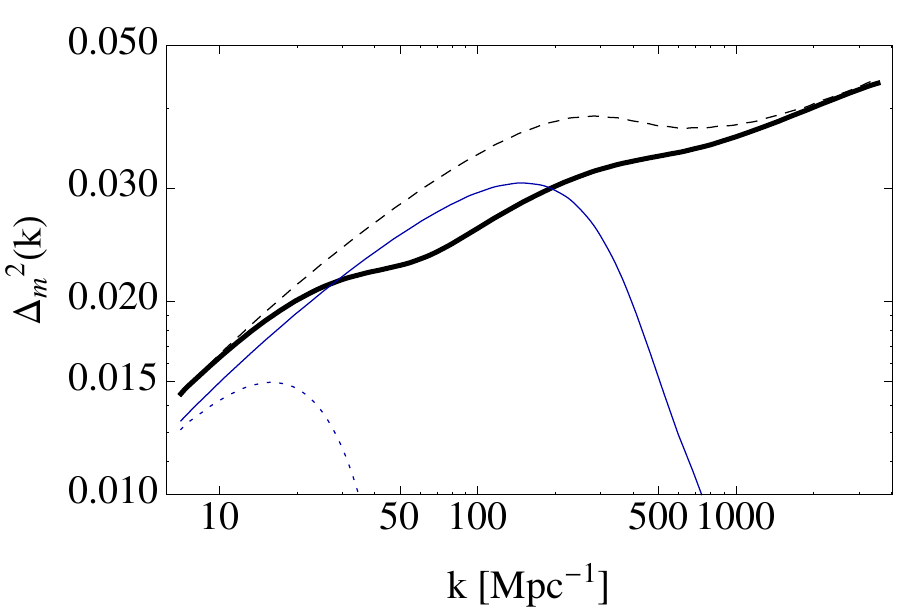}
\caption{  IR resummation of the total matter power spectrum at $z=40$.  The dashed black line is the total linear power spectrum, and the solid black line is the tree-level IR resummation.  The solid blue curve shows the tree-level cross correlation $\la \delta^{(1)}_b \delta^{(1)}_c \ra$ contribution to the total power spectrum, and the dotted blue line shows the same contribution but after IR resummation.  The suppression of this contribution is the reason for the advection effect.      }\label{hirataplot}
\end{center}
\end{figure}

We should comment briefly about our choice of $\Lambda_{\rm Resum}$.  We need to choose it large enough so that we resum most of the IR displacements, but it has to be small enough so that we do not include large $\delta$ fluctuations that might require some counterterm.  Thus, we will chose it just above the scale that includes most of the IR displacements.  To find that scale, consider the size of the displacements as a function of the cutoff $\Lambda_{\rm Resum}$:
\be\label{eq:lambdaIR}
(\delta^\sigma_{s <})^2 = \int_0^{\Lambda_{\rm Resum}} \frac{ d^3 k }{ (2 \pi )^3 } \frac{P^\sigma_{ 11} }{k^2 } \ . 
\ee
We should choose $\Lambda_{\rm Resum}$ when the above integral saturates, which we find is at $\Lambda_{\rm Resum} = 2  \rm Mpc^{-1}$. A plot of (\ref{eq:lambdaIR}) as a function of $\Lambda_{\rm Resum}$ is given in Fig.~\ref{fig:lambdaresum}.
\begin{figure}[htb!]
\centering
\begin{tabular}{cc}
\includegraphics[width=8cm]{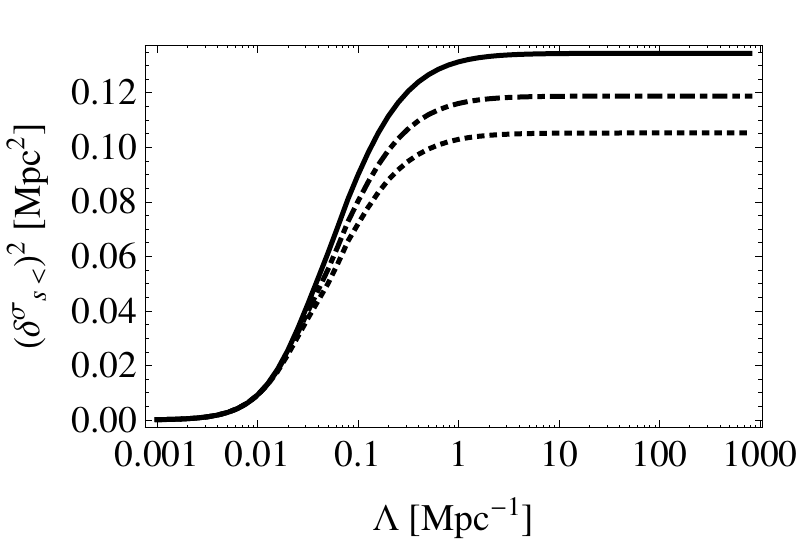}  
\end{tabular} 
\caption{Equation~(\ref{eq:lambdaIR}) as a function of $\Lambda=\Lambda_{\rm Resum}$ for $\sigma=c,b,bc$. The solid line is the curve for dark matter, the dotted line is the curve for baryons, and the dot-dashed curve is the one for the cross correlation. \label{fig:lambdaresum}}
\end{figure}

%
%
%
%

\section{Comparison to Nonlinear Data\label{data}}

Here we compare our analytic results to the non-linear simulation data that includes baryonic effects of~\cite{vanDaalen:2011xb, Schaye:2009bt}.  We have at our disposal a total of twelve different simulations, each with different baryonic physics and/or cosmological parameters included.  The names of these simulations are given in Table \ref{simnames}~\footnote{For more details on the baryon physics employed in these simulations, see \cite{Schaye:2009bt,Wiersma:2008cs,stellar,Schaye:2007ss,Vecchia:2008kn,AGN}. Implications for cosmic shear studies are discussed in \cite{shear1,shear2}, and the model {\it AGN} is compared to X-ray and optical observations of groups and clusters in \cite{McCarthy:2009kk,Brun:2013yva}.  For example, these studies showed that the simulations that included AGN feedback matched observations of stellar mass fractions, star formation rates, and stellar age distribution much better than simulations that did not include AGN feedback. }.    In our calculations, we will use the one-loop IR-resummed quantities.  Since in this case we are including the one-loop term, there is a subtlety.  We have neglected the isocurvature terms in the loops because they are negligible there. However, we have kept them at tree level. This means that the one-loop implementation of the IR resummation on the linear power spectrum needs to be slightly modified from the form given in \cite{Senatore:2014via}.  We accomplish this by using
\begin{align}
P^\alpha(k ; t) |_1 =  \int dk' \bigg(  \hat{M}^\alpha_{||_{1}} ( k , k' ; t , t) P^\alpha_{ 11} ( k'; t)   + \hat{M}^\alpha_{||_{0}} ( k , k' ; t , t ) P^A_{ 1-loop} ( k'; t)  \bigg), 
\end{align}
where $\alpha = c , b , bc$, and $P^A_{\rm 1-loop}$ is the loop calculated with only the adiabatic mode.  The modification is to use $K_0^\alpha[X_\alpha , Y_\alpha] ( \kvec, \qvec ; t_1 , t_2) \cdot {(K_0^\alpha)}^{-1}[X_A,Y_A] ( \kvec , \qvec ; t_1 , t_2 ) \Big| \Big|_{1}$ in the expression $\hat{M}^\alpha_{||_{1}} ( k , k' ; t , t)$.  Here, we have explicitly displayed the functional dependence of $K_0$ on the $X$ and $Y$ functions.  The reason for this is that the factor of $(K_0^\alpha)^{-1} ( \kvec , \qvec ; t_1 , t_2 ) \Big| \Big|_{1}$ ensures that we do not double-count corrections in $\esl$, which have been fully non-perturvatively resummed in the factor of $K_0^\alpha[X_\alpha , Y_\alpha] ( \kvec, \qvec ; t_1 , t_2)$, by canceling the piece proportional to $\esl$ that comes in $P_{\rm 1-loop}$. Said another way, if we were to use the naive ${(K_0^\alpha)}^{-1}[X_\alpha,Y_\alpha] ( \kvec , \qvec ; t_1 , t_2 ) \Big| \Big|_{1}$, this would incorrectly try to compensate for $\esl^{I}$ in the one-loop term ($\esl^{I}$ is defined as in (\ref{eq:epsparameters}) but with $P_{11}^{I}$ in the integral), but since these were not included in $P_{\rm 1-loop}^A$, we do not need to compensate for them.  
\begin{table}
\begin{tabular}{| l | p{11.7cm} |}
\hline
  Simulation & Description  \\  \hline
  $AGN$ & Includes AGN (in addition to SN feedback)  \\
  $AGN\_WMAP7$ & Same as AGN, but with a WMAP7 Cosmology \\
  $DBLIMFV1618 $  &   Top-heavy IMF at high pressure, extra SN energy wind velocity \\
  $DMONLY$ & No baryons (cold dark matter only) \\
  $DMONLY\_ WMAP7$   &  Same as DMONLY, but with a WMAP7 cosmology\\
  $NOSN $    & No SN energy feedback \\
  $NOSN \_ NOZCOOL $& No SN energy feedback and cooling assumes primordial abundances  \\
  $NOZCOOL $ & Cooling assumes primordial abundances \\
  $WDENS $ & Wind mass loading and velocity depend on gas density (same SN energy as REF) \\
  $WML1V848 $& Wind mass loading $\eta = 1$, velocity $v_w = 848 \rm km/s$  (same SN energy as REF)   \\
  $WML4$ & Wind mass loading $\eta = 4$ (twice the SN energy as REF) \\ 
  $REF$ &  Reference simulation \\\hline
\end{tabular}  
\caption{  Description of simulations done in \cite{vanDaalen:2011xb}. }  \label{simnames} 
\end{table}
  
  %
  %
  %
  %

 \subsection{Effect of Baryons on Total Matter Power Spectrum} \label{baryonsontotal}
 
 We will now compare our calculations to the results of numerical simulations that include baryon physics. As we already described, we will use data from the simulations discussed in  \cite{vanDaalen:2011xb}, which are summarized in Table \ref{simnames}. According to the authors, the most up-to-date data uses WMAP7 parameters $ \Omega_m=0.272 , \ \Omega_b= 0.0455 , \  \Omega_\Lambda=0.728 , \ \sigma_8~=0.81 , \ n_s =0.967, \ h=0.704  $, and feedback from active galactic nuclei~(AGN).  Here we compare the effect of baryons on the total matter power spectrum (also known as the adiabatic power spectrum) with the dark-matter only simulation, i.e. we will be concerned with the quantity 
 \be\label{eq:ra}
 R(k) \equiv\frac{ P^A_{\rm with \, baryon} (k) }{ P^A_{\rm DM \, only}(k)}\ .
 \ee 
 This is a useful ratio because most of the cosmic variance cancels in the simulation data, and only other small systematic errors remain.  To address those errors, for which we do not have specific information, we add a $1\%$ systematic error in quadrature to our theoretical errors in all of our plots.  In order to study $R$, we parameterize the adiabatic speed of sound as 
\be
\bar{c}_A^2 ( a_0 ) =   \ccs+  w_b \, \dca+ \mathcal{O}(w_b^2)  \ ,
\ee
where we expect $ \dca $ to be of order one in units of $1/\knl^2$, and $\ccs$ is the speed of sound in a dark-matter only universe, which is the $w_b \rightarrow 0 $ limit of the expression in \eqn{csai}.

In our notation, the ratio is given by
\be
R_{\rm EFT}(k) = \frac{ P^A( k ; w_b , \ccs , \dca) }{ P^A( k ; 0 , \ccs , 0 )} \ ,
\ee  
where we have written the explicit dependence on the baryon fraction $w_b$, the dark-matter-only speed of sound, $ \ccs$, and the deviation $\dca$.  We first determine the dark-matter-only speed of sound $ \ccs$ by fitting to the non-linear data without baryons.  Below we plot the result, where we have determined the speed of sound to be $ \ccs \simeq 7.9 \,\knl^{-2}$.  Once we have determined $ \ccs$, we use the baryon simulation to determine $\dca$.  We can then do the same for the set of simulations that use the WMAP3 cosmology  $\Omega_m= 0.238,\ \Omega_b= 0.0418,\ \Omega_\Lambda=  0.762, \ \sigma_8 = 0.74, \ n_s =  0.951 , \ h= 0.73 $.   There, we determine that $\ccs \simeq 9.6 \,\knl^{-2}$~\footnote{These values of  $c_A$ are a bit larger than order one. Notice however that their size is very sensitive to the definition of $\knl$: it is enough to change $\knl$ by a factor of 2, to decrease the values of $c_A$ by a factor of 4.}.  We plot the results in Figs.~\ref{determinecs}.

\begin{figure}[htb!]
\centering
\begin{tabular}{cc}
\includegraphics[width=8cm]{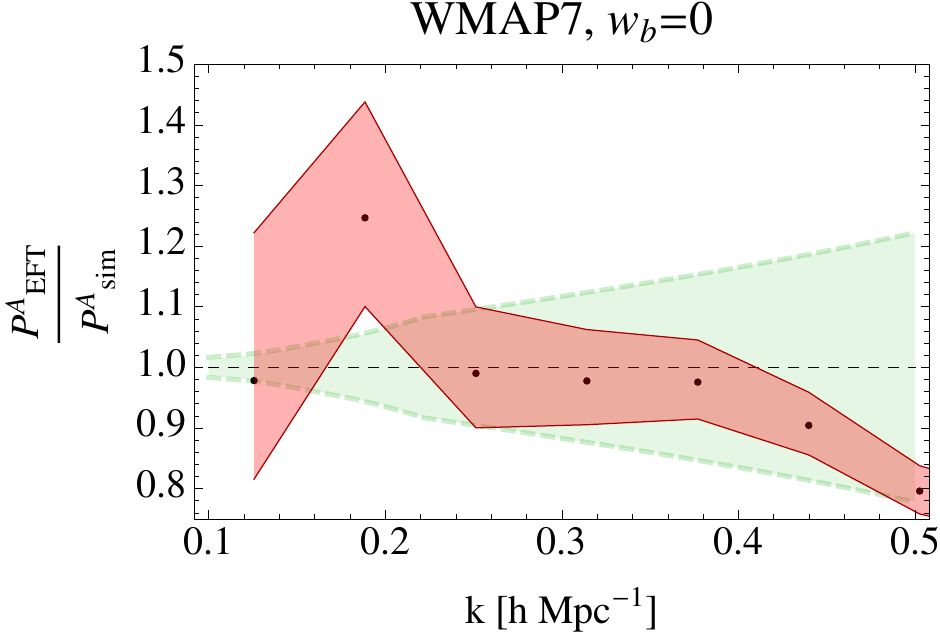}  & \includegraphics[width=8cm]{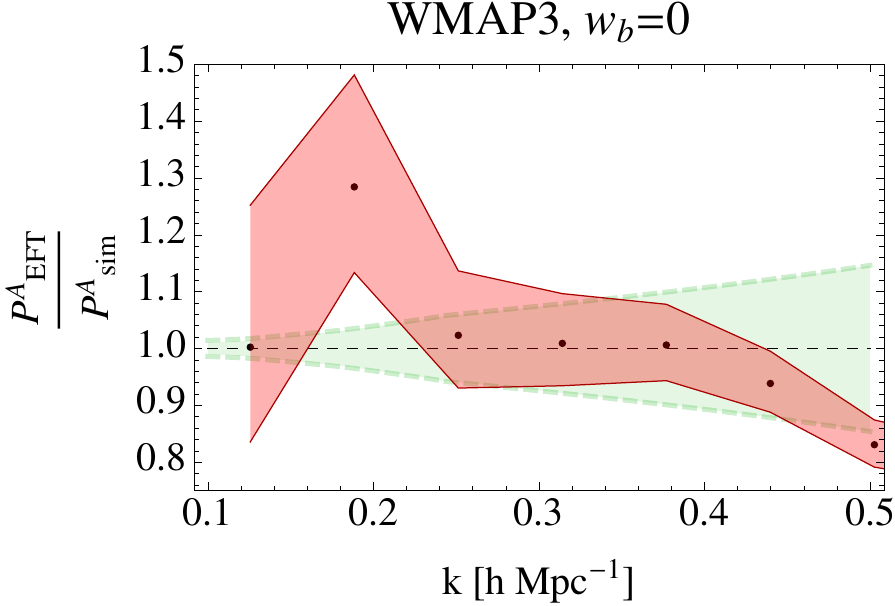}
\end{tabular} 
\caption{ We use the dark-matter-only simulation to determine the speed of sound $ \ccs \simeq 7.9 \,\knl^{-2}$ for the WMAP7 data, and $ \ccs \simeq 9.6 \,\knl^{-2}$ for the WMAP3 data.  We do this by plotting the ratio of the EFT total matter power spectrum with $w_b = 0$, $P^A ( k ; 0 , \ccs , 0)$, to the simulation data for the same quantity.  The black dots are the data points, and the surrounding red region is the error due to the cosmic variance of a box size $L = 100 \, h^{-1} \rm Mpc$.  The green region is the size of the theoretical error, which we have calculated by estimating the size of the two-loop corrections that we have not included, using \eqn{estimate11}, and the green dashed line is this error added in quadrature with a $1\%$ error for unknown systematics.} \label{determinecs}
\end{figure}

\begin{figure}[htb!] 
\centering
\begin{tabular}{cc}
\includegraphics[width=8.2cm]{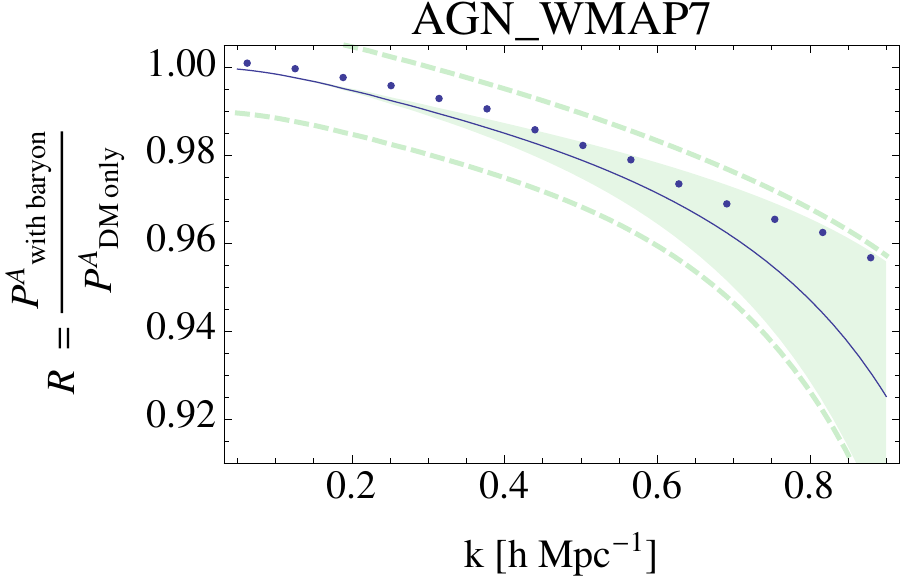} & \includegraphics[width=8.cm]{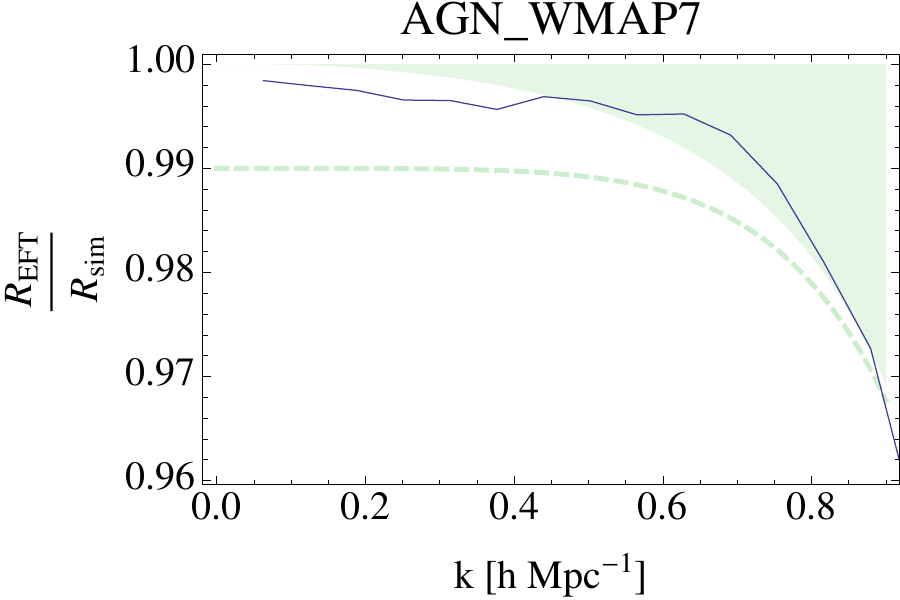}
\end{tabular}
\caption{  We fit to the simulation that includes baryons and determine $\dca \simeq 1.32 \,\knl^{-2}$.  In these plots, we compare the ratio of the adiabatic power spectra in the presence of baryon effects and in their absence, $R = P^A_{\rm with \, baryon} / P^A_{\rm DM\, only}$, as calculated in the EFT to the same quantity calculated from the data.  The solid line in the left panel is $R_{\rm EFT}$, and the points are from the simulation data.  The fit starts deviating near $k \approx 0.7  \ h \rm Mpc^{-1}$.  The green region is the size of the theoretical error, which we have calculated by estimating the size of the two loop corrections that we have not included, using Eqs. (\ref{estimatedelta1}) and (\ref{estimatedelta2}). The dashed line is the same theoretical error after adding in quadrature a 1\% error for unknown systematics. }  \label{agnwmap7ratio}
\end{figure}

\begin{figure}[htb!] 
\begin{center}
\includegraphics[width=14cm]{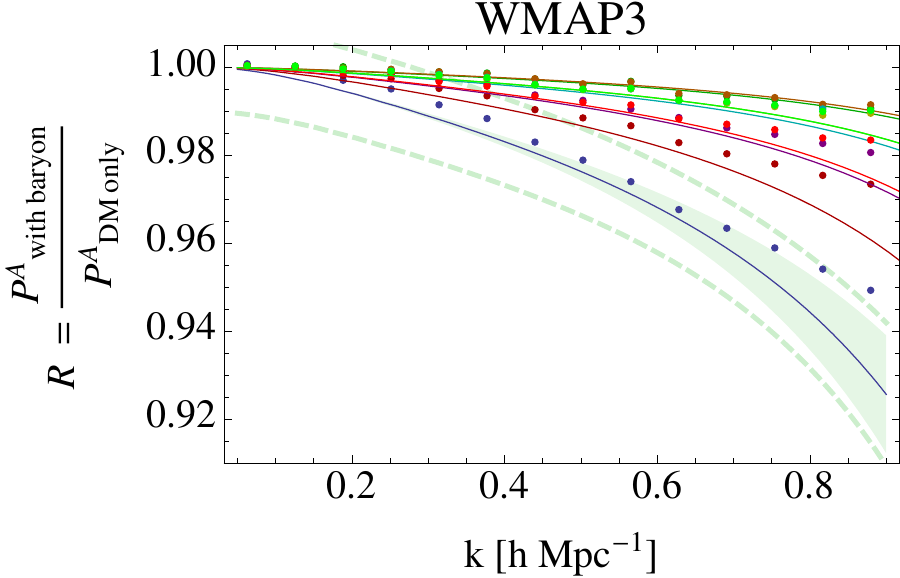}
\caption{  We fit to the simulations that include various baryonic effects by comparing  the ratio of the adiabatic power spectra in the presence of baryon effects and in their absence, $R = P^A_{\rm with \, baryon} / P^A_{\rm DM \, only}$, as calculated in the EFT to the same quantity calculated from the data.  Each simulation has a different best-fit value of $\dca$.  Here, we obtain a range of $\dca$: $\dca \simeq 2.4 \,\knl^{-2}$ is the blue curve, which is the $AGN $ data, while $\dca \simeq 0.34 \,\knl^{-2}$ is the yellow curve, which is the $NOSN\_ NOZCOOL$ simulation.  The rest of the curves are $DMBLIMFV1618$ (dark red), $NOSN$ (dark green), $NOZCOOL$ (cyan), $REF$ (dark yellow), $WDENS$ (purple), $WML1V848$ (red), $WML4$ (green).  The green region is the size of the theoretical error, which we have calculated by estimating the size of the two loop corrections that we have not included, using Eqs. (\ref{estimatedelta1}) and (\ref{estimatedelta2}).  The dashed line is the same theoretical error after adding in quadrature a 1\% error for unknown systematics. This has only been plotted for the $AGN$ simulation to avoid clutter.}  \label{wmap3_fit}
\end{center}
\end{figure}

\begin{figure}[htb!] 
\begin{center}
\includegraphics[width=14cm]{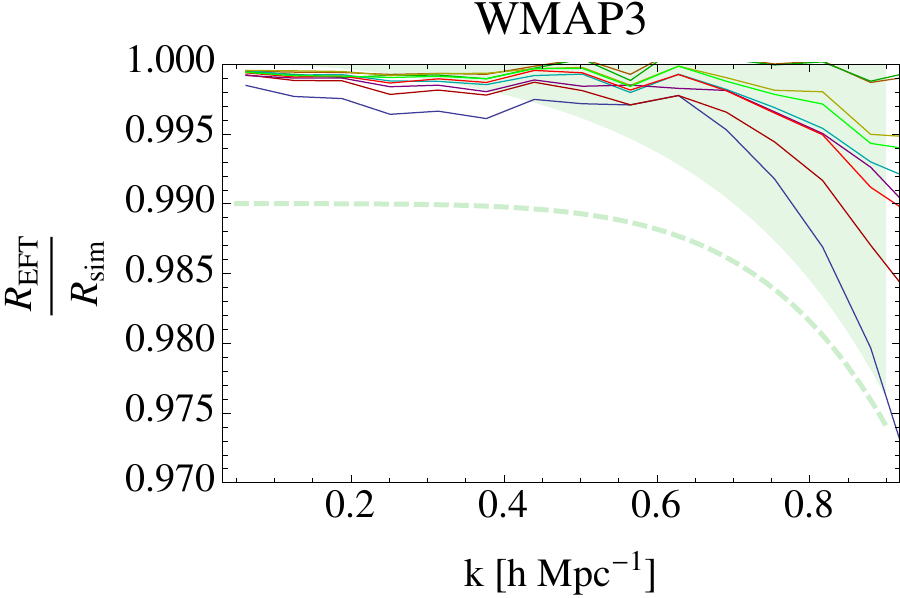}
\caption{    We plot the relative difference between our EFT predictions and the baryon simulations for  the ratio of the adiabatic power spectra in the presence of baryon effects and in their absence, $R = P^A_{\rm with \, baryon} / P^A_{\rm DM \, only}$.  The color coding is the same as in Figure \ref{wmap3_fit}.  The green region is the size of the theoretical error, which we have calculated by estimating the size of the two loop corrections that we have not included, using Eqs. (\ref{estimatedelta1}) and (\ref{estimatedelta2}).  The dashed line is the same theoretical error after adding in quadrature a 1\% error for unknown systematics.   }  \label{wmap3ratio}
\end{center}
\end{figure}

Notice that the fit to both dark-matter-only simulations fails near $ k =  0.4 \, h \rm Mpc^{-1}$, consistent with expectations coming from the size of two-loop terms in the EFT.  The theoretical error in Fig.~\ref{determinecs} (as well as in Fig.~\ref{pbratios} below) is due to the two-loop contribution that we do not calculate.  To estimate this error, we first express the linear power spectrum as a piecewise power law~\cite{Carrasco:2013mua,Pajer:2013jj}
\be
P^A_{11, \rm pl}(k) =  (2\pi)^3 \begin{cases}  \frac{1}{ \knl^3 } \left( \frac{ k}{  \knl } \right)^{n} & \text{for} \, \,  \, k> k_{\rm tr}  \\    \frac{1}{ \tilde{k}_{\rm NL}^3 } \left( \frac{ k}{  \tilde{k}_{\rm NL} } \right)^{\tilde{n}}  & \text{for} \, \, \, k < k_{\rm tr}   \end{cases}
\ee
where $k_{\rm tr}$ is the transition scale between the two power laws.  For the WMAP7 data, we find $\knl = 4.3 \unitsk$, $\tilde{k}_{\rm NL} = 1.73 \unitsk$, $n = -2.1$, $\tilde{n} = -1.7$, and $k_{\rm tr} =0.22 \unitsk$.  For the WMAP3 data, we find $\knl = 5.50 \unitsk$, $\tilde{k}_{\rm NL} = 2.68 \unitsk$, $n = -2.1$, $\tilde{n} = -1.83$, and $k_{\rm tr} =0.24 \unitsk$.  The larger value of $\knl$ in the WMAP3 data is due to the smaller value of $\sigma_8$.  By using the scaling relation $P^A_{\rm L-loop} / P^A_{11} \sim ( k / \knl )^{L(3 + n)}$ where $L$ is the loop order, the leading piece that we did not calculate is from $P^A_{\rm 2-loop}$  and takes the form
\be \label{estimate11}
\Delta P^{A}_{\rm EFT}(k)  = \alpha \, (2 \pi)^2  P^A_{11,\rm pl}(k)  \begin{cases}   \left( \frac{k}{\knl} \right)^{2 (3+n)}   & \text{for} \, \, \, k>k_{\rm tr} \\    \left( \frac{k}{\tilde{k}_{\rm NL}} \right)^{2 ( 3 + \tilde{n})}  & \text{for} \, \, \, k < k_{\rm tr}  \ .   \end{cases} 
\ee
Here $\alpha$ is an order one number which attempts to account for combinatoric factors in various two-loop diagrams.  We have used $\alpha = 1/2$~\cite{Carrasco:2013mua} in Figs. \ref{determinecs} and \ref{pbratios}.  

In Figs. \ref{agnwmap7ratio}, \ref{wmap3_fit}, and \ref{wmap3ratio}, we plot the comparison of the predictions of the EFT with simulations for the quantity $R(k)$. We naively would expect the EFT to fail at $k \approx  0 . 4 \, h \rm Mpc^{-1}$ because this is when the two-loop terms become important. However, we observe that the EFT with baryons matches the quantity $R(k)$ out to about $k \approx 0 . 7 \, h \rm Mpc^{-1}$.  In reality, this is still consistent with the expectation that the one-loop computation matches to   $ k \sim 0.4  \, h \rm Mpc^{-1}$ because, in the ratio that we consider, higher order terms that we do not compute come with an extra factor of $w_b$.  To see this, we write the ratio as:
\be \label{newton}
\frac{ P^A_{\rm with \, baryon} }{P^A_{\rm DM \, only} } = 1 - \frac{ 2  (2\pi) w_b  \, \dca \, k^2 P^A_{11}  }{ P^A_{11} + P^A_{\rm 1-loop} - 2 (2\pi) \ccs \, k^2 P^A_{11}  } + \Delta ( k ) \ . 
\ee
We are interested in calculating this ratio to within a few percent.  It is straightforward to estimate that the cosmic variance of ratios such as (\ref{newton}) is below this level, so we ignore it for this kind of quantities. Concerning the theoretical error, the leading terms that we did not calculate in (\ref{newton}) are  
\be
\Delta(k) \supset \frac{1}{P^A_{\rm DM \, only} |_1  } \left( P^I_{\rm 1-loop} + k^4  (2\pi)^2\left( 2 w_b \, \ccs \, \dca  + ( \dca )^2 w_b^2 \right) P^A_{11} +  w_b \,  (2\pi) \dca  k^2 P^A_{\rm 1-loop} \right) \ ,
\ee
where $P^I_{\rm 1-loop} $ is the contribution from all the one-loop diagrams that have one isocurvature mode, and $P^A_{\rm DM \, only} |_1$  is the IR resummed adiabatic power spectrum \emph{up to} one loop.  $P^I_{\rm 1-loop} \sim w_b 10^{-2} P^A_{\rm 1-loop}$, and so is subleading with respect to $(2\pi) w_b \dca k^2 P^A_{\rm 1-loop} $ for $k~\gtrsim 0.1 \, h \rm Mpc^{-1}$.  This can be seen because $\delta_c = \delta_A + w_b \delta_I$, so the difference between the CDM one-loop term and the adiabatic one-loop term is 
\be
\langle \delta_c^2 \rangle_{\rm 1-loop} - \langle \delta_A^2 \rangle_{\rm 1-loop} = 2 w_b \langle \delta_A^{(3)} \delta_I^{(1)} \rangle + \dots\ , 
\ee
and this term is down by a factor of $10^{-2} w_b$ from the adiabatic one-loop term $\langle \delta_A^{(3)} \delta_A^{(1)} \rangle$.  The $\dots$ in the above equation represents terms with higher powers of $w_b$, or with higher order isocurvature modes.  There are two relevant contributions to $\Delta(k)$, 
\begin{align}
\Delta_1 &= \alpha_1 \, w_b(2\pi) \left( \frac{k }{\knl} \right)^2 \frac{P^A_{\rm 1-loop}}{P^A_{\rm DM \, only}  |_1 }  \n
& = \alpha_1 \, w_b(2\pi) \left( \frac{k }{\knl} \right)^2  (2\pi)    \frac{P^A_{11,\rm pl}}{P^A_{\rm DM \, only}|_1}    \begin{cases}   \left( \frac{k}{\knl} \right)^{3 + n} & \text{for} \, \, \, k > k_{\rm tr} \\   \left( \frac{k}{\tilde{k}_{\rm NL}} \right)^{3 + \tilde{n}} & \text{for} \, \, \, k < k_{\rm tr}  \  ,  \end{cases}  \label{estimatedelta1}
\end{align}
and
\be
\Delta_2 =  w_b \, (2 \pi)^2 \left( \frac{k }{\knl} \right)^4 \frac{P^A_{11,\rm pl}}{P^A_{\rm DM\,  only} |_1} \ .  \label{estimatedelta2}
\ee 
We have plotted the theoretical error due to these contributions in Figs. \ref{agnwmap7ratio}, \ref{wmap3_fit}, and \ref{wmap3ratio}.  Again, $\alpha_1$ gives an estimate of the combinatoric factors, and we have used $\alpha_1 = 1/2$ as before. We find that $\Delta_1(k) + \Delta_2 (k) \approx 0.02$ at $k \approx 0.8 \unitsk$.  This is consistent with the point at which the theory stops fitting the data.

We can do this estimate in a different way by comparing to the previous estimates in the one-fluid case. We know that the one-loop dark matter power spectrum itself fails at $k_* \approx 0.4\, h \rm Mpc^{-1}$ due to the same terms we are trying to estimate, which differ simply by a factor of $w_b$. Therefore, to obtain when the theoretical error becomes important, we just have to rescale $k_*$ to find where our calculation should fail.  Thus we expect $R(k)$ to deviate by 3\% at $k \approx k_* / {w_b}^{1/2.9} \approx  0.74 \, h {\rm Mpc}^{-1}$, which is consistent with the estimates from (\ref{estimatedelta1}) and (\ref{estimatedelta2}).   

Overall,  Figs. \ref{agnwmap7ratio}, \ref{wmap3_fit}, and \ref{wmap3ratio} seem to show that the EFT is able to describe a large range of baryonic processes for the total power spectrum.  For $k \lesssim 0.6 \, h \rm Mpc^{-1}$, we see that all of these processes are well described by Eq.~(\ref{newton}), which shows that the functional dependence of the effect of baryons is of the form $k^2 P_{11}$, as predicted by the EFT, and the only parameter is $\dca$.  All of the particulars of the baryonic physics are describable, at long distances, by the particular value of $\dca$, which measures how different the adiabatic speed of sound is from the single-species, dark-matter-only speed of sound.  Thus, we have a definite functional form for the effect of baryons on the total matter power spectrum. Furthermore, we find $\dca$ is of order unity in units $\knl^{-2}$, suggesting that the scale at which the baryons become non-linear is comparable to the one of dark matter.

%
%
%
%

\subsection{Baryon Power Spectrum}
In this section, we use the same simulations to compare the predictions of the EFT for the baryon power spectrum itself.  Using the values of $\ccs$ and $\dca$ determined above, we can then fit to the baryon power spectrum $P^b$ to determine $\ci$.  To do this, we consider the ratio of the baryon power spectrum to the total power spectrum
\be\label{eq:ratiob}
R^b ( k ) \equiv \frac{ P^b (k) }{P^A (k)} \ . 
\ee
The results are presented in Figs. \ref{wmap7ratiobar}, \ref{wmap3_barfit} and \ref{wmap3_barfitratio}.  For the WMAP7 cosmology simulation, we determine that $\ci \simeq-1.18 \,\knl^{-2}$.  As an indication, using (\ref{eq:simple_matching}), we find $c_c^2\simeq 9$ and $c_b^2\simeq 10$. Notice that, as intuitive, the star-formation physics makes the speed of sound of baryons larger, but not by much, than the gravitational one. Instead, for the various WMAP3 cosmology simulations (each of which includes different baryonic physics), we determine $\ci$ in a range $-2.17 \,\knl^{-2}$ ($AGN$) to $-0.72 \,\knl^{-2}$ ($NOSN\_ NOZCOOL$).  All of the fits are within about $2\%$ of the data up to $k \approx 0.6 \unitsk$.  In Fig. \ref{pbratios} we plot $P^b_{EFT} / P^b_{\rm sim}$ for the $AGN \_ WMAP7$  and the $AGN$ simulations with the values of $\ci$ determined using the $R^b(k)$ fits. Consistently, we find that the fits are well within the cosmic variance up to when the theoretical error becomes sizable.    

\begin{figure}[htb!] 
\centering
\begin{tabular}{cc}
\includegraphics[scale=.86]{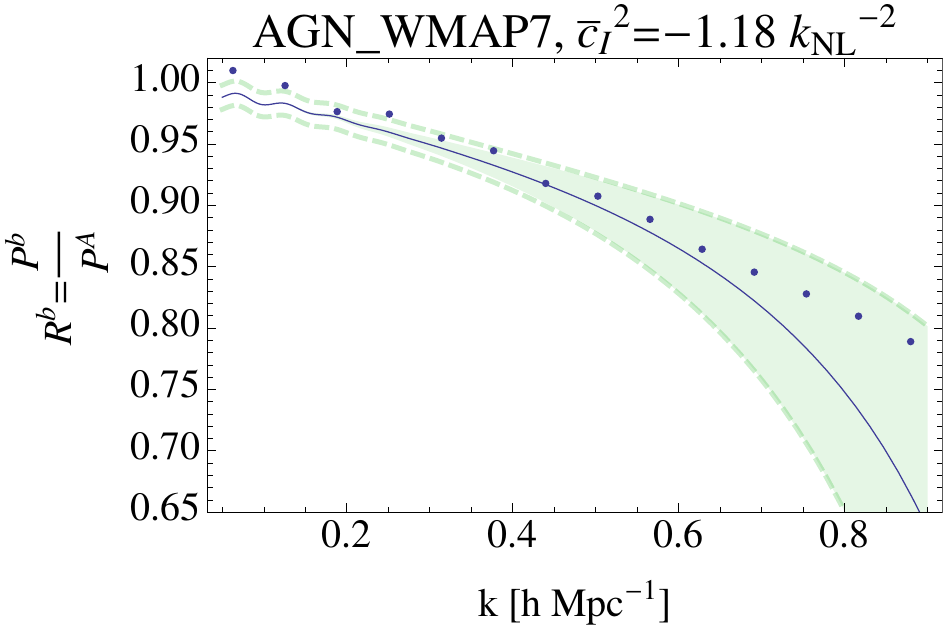} & \includegraphics[scale=.82]{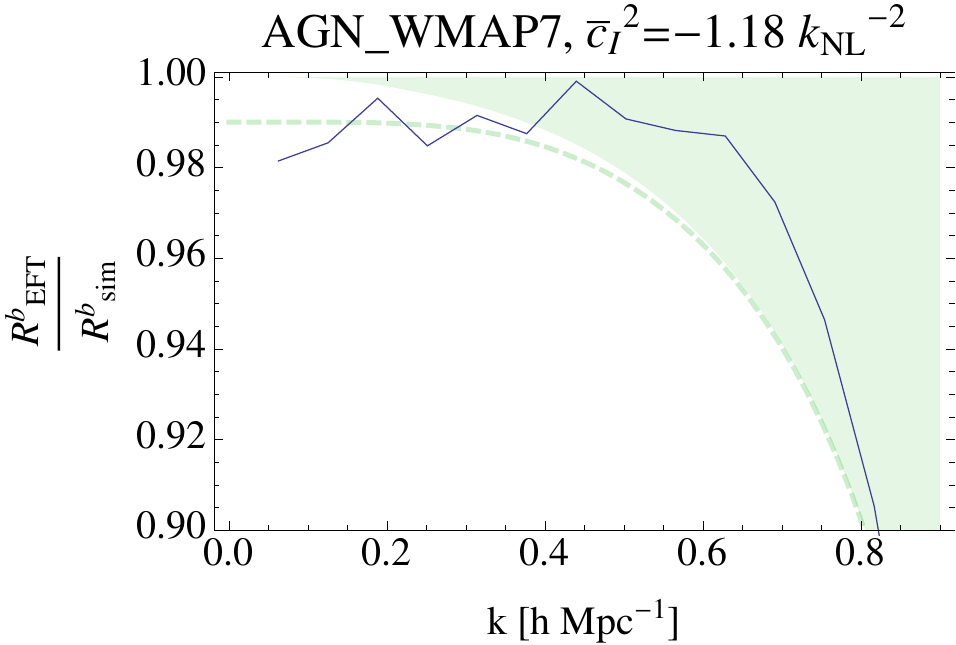} 
\end{tabular}
\caption{We plot  the ratio of the baryon power spectrum and the adiabatic power spectrum, $R^b = P^b / P^A$, as calculated in the EFT to the same quantity calculated from the data, and determine $\ci \simeq-1.18 \,\knl^{-2}$ by choosing the one that better matches the data until the theoretical errors become relevant. The solid line in the left panel is $R^b_{\rm EFT}(k)$, and the points are from the simulation data.  The fit starts deviating near $k \approx 0.7 \ h \rm Mpc^{-1}$.  This is consistent with the failure of the EFT at $k \approx 0.4 \unitsk$ because here we are comparing the ratio $R^b \equiv P^b/P^A$, and some higher order terms that we do not compute cancel in the ratio.  The green shaded region is the size of the theoretical error, which we have calculated by estimating the size of the two loop corrections that we have not included, using Eqs.~(\ref{deltab1}) and~(\ref{deltab2}), and the green dashed line is this error added in quadrature with a $1\%$ error for unknown systematics.}  \label{wmap7ratiobar}
\end{figure}

As in Section \ref{baryonsontotal}, the failure of the $R^b$ fit at $k \approx 0.7 \unitsk$ (Figs. \ref{wmap7ratiobar}, \ref{wmap3_barfit}, and \ref{wmap3_barfitratio}) is consistent with the failure of the prediction for the power spectrum at $k \approx 0.4 \unitsk$ (Fig. \ref{pbratios}).  Consider the expression
\be \label{reimann}
\frac{ P^b }{P^A } = \left( \frac{P^b}{P^A} \right)_{\rm 1-loop} + \Delta^b ( k )  
\ee
where $\left( P^b/P^A \right)_{\rm 1-loop} $ is the part of the ratio that we have calculated in the EFT up to one-loop, and $\Delta^b(k)$ are the terms that we have not calculated.  The leading terms in  $\Delta^b(k)$ are
\be
\Delta^b (k) \supset \frac{1}{P^A |_1} \left( P^{b,I}_{\rm 1-loop} + k^4  (2\pi)^2 \left( -2 w_c \, \ca \, \ci +  w_c^2 (\ci)^2 \right) P^A_{11} +   (2\pi) w_c \, \ci  k^2 P^A_{\rm 1-loop} \right) \ ,
\ee
where $P^{b,I}_{\rm 1-loop}$ is the one-loop term in the baryon power spectrum with one isocurvature mode inserted.  This term is of order $P^{b,I}_{\rm 1-loop} \sim  w_c 10^{-2} P^A_{\rm 1-loop}$, and is comparable to $(2\pi)w_c \, \ci  k^2 P^A_{\rm 1-loop} $ near $k \sim 0.1$ $\unitsk$, but less important at higher $k$. Therefore, as before, there are only two relevant contributions to~$\Delta^b(k)$, 
\begin{align}
\Delta^b_1 & = \alpha_1^b \, w_c(2\pi) \left( \frac{k }{\knl} \right)^2 \frac{P^A_{\rm 1-loop}}{P^A_{\rm DM only}  |_1 }  \\
& = \alpha_1^b \, w_c(2\pi) \left( \frac{k }{\knl} \right)^2  (2\pi)    \frac{P^A_{11,\rm pl}}{P^A |_1}    \begin{cases}   \left( \frac{k}{\knl} \right)^{3 + n} & \text{for} \, \, \, k > k_{\rm tr} \\   \left( \frac{k}{\tilde{k}_{\rm NL}} \right)^{3 + \tilde{n}} & \text{for} \, \, \, k < k_{\rm tr}  \  ,  \end{cases}  \label{deltab1}
\end{align}
and
\be
\Delta^b_2 =  w_c \,  (2 \pi)^2 \left( \frac{k }{\knl} \right)^4 \frac{P^A_{11}}{P^A |_1} \ . \label{deltab2}
\ee 
Again, $\alpha^b_1$ gives an estimate of the combinatorics, and we have taken it to be again $\alpha_1^b = 1/2$.  We have plotted the theoretical error due to these contributions in Figs. \ref{wmap7ratiobar}, \ref{wmap3_barfit}, and \ref{wmap3_barfitratio}.  We find that $\Delta^b_1(k) + \Delta^b_2 (k) \approx 0.02$ and $k \sim 0.55 \unitsk$ for the WMAP7 data.  This is also consistent with the estimate we get from knowledge of when the theory fails for the power spectrum.  Since the one-loop dark matter power spectrum itself fails at $k_* \approx 0.4\, h \rm Mpc^{-1}$ due to a two-loop term that is present also here, just multiplied by $w_c \ci/\bar c_A^2$, we just have to rescale $k_*$ to find where our calculation should fail, and we find that $R(k)$ deviates by 3\% at $k \approx k_* / ( w_c\, \ci/ \bar c_A^2)^{1/2.9} \approx 0.9\, h {\rm Mpc}^{-1}$.  In this estimate we include the factor of $\ci$ because it is quite smaller than $\bar c_A^2$. This estimate is in comfortable agreement with the former one.

\begin{figure}[htb!] 
\begin{center}
\includegraphics[width=14cm]{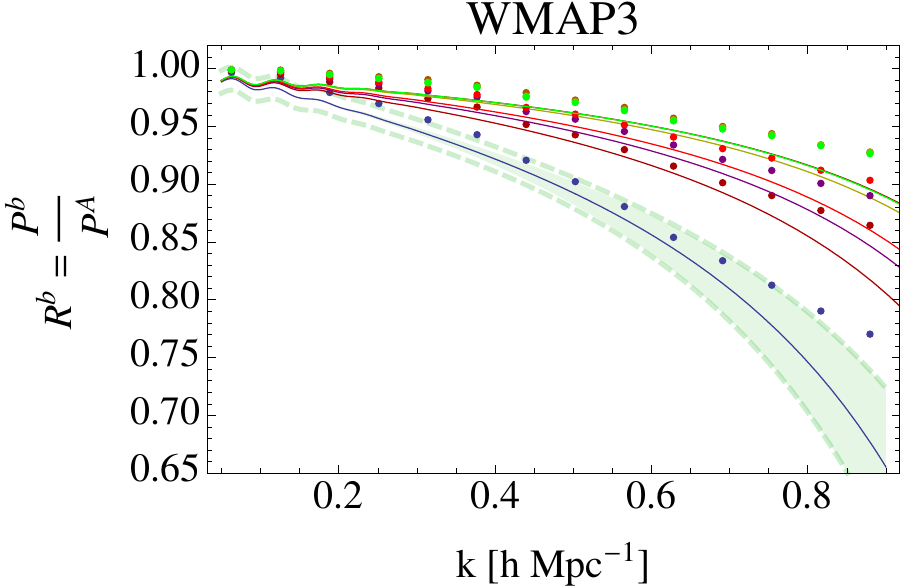}
\caption {We fit the baryon power spectrum in simulations that include various baryonic effects.  We choose to make the fit by comparing the ratio of the baryon power spectrum and the adiabatic power spectrum, $R^b = P^b / P^A$, that is  as calculated in the EFT to the same quantity calculated from the data.  Each simulation has a different best-fit value of $\ci$.  Here, we obtain a range of $\ci$:  $\ci \simeq -2.17 \,\knl^{-2}$ is the blue curve, which is the $AGN $ data, while $\ci \simeq -0.72 \,\knl^{-2}$ is the yellow curve, which is the $NOSN\_ NOZCOOL$ simulation.  The rest of the curves are $DMBLIMFV1618$ (dark red), $REF$ (dark yellow), $WDENS$ (purple), $WML1V848$ (red), $WML4$ (green).  The green shaded region is the size of the theoretical error, which we have calculated by estimating the size of the two loop corrections that we have not included, using Eqs. (\ref{deltab1}) and (\ref{deltab2}), and the green dashed line is this error added in quadrature with a $1\%$ error for unknown systematics. This has only been plotted for the $AGN$ simulation to avoid clutter.}

 \label{wmap3_barfit}
\end{center}
\end{figure}

\begin{figure}[htb!] 
\begin{center}
\includegraphics[width=14cm]{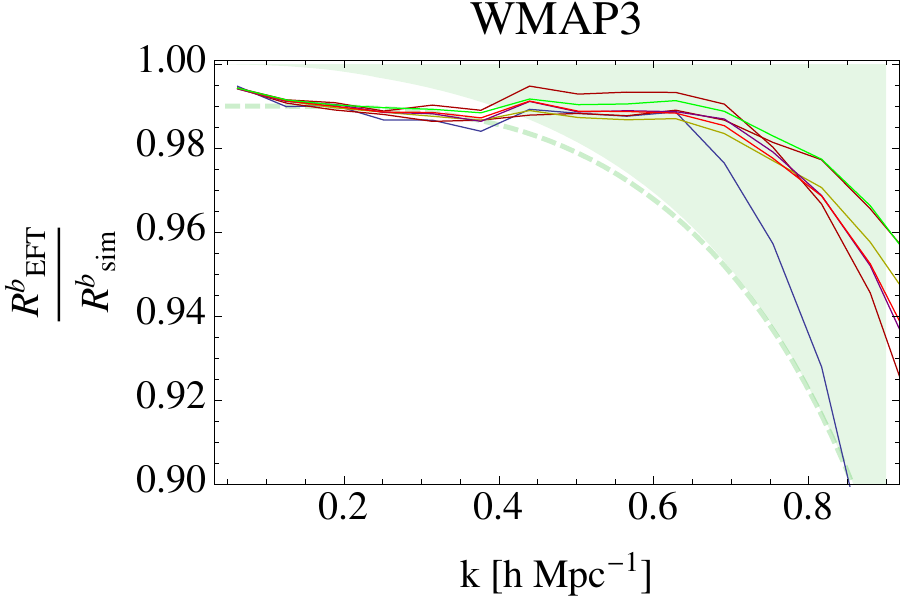}
\caption{We plot the relative difference between our EFT calculations and the baryon simulations.  We choose to make the fit by comparing the ratio of the baryon power spectrum and the adiabatic power spectrum, $R^b = P^b / P^A$, as calculated in the EFT to the same quantity calculated from the data.  The color coding is the same as in Figure \ref{wmap3_barfit}.  These fits all start to fail near $k \approx 0.7 \unitsk$, even though the EFT starts failing in the power spectrum at one loop at $k \approx 0.4 \unitsk$, as seen in Fig. \ref{pbratios}.  This is consistent because here we are comparing the ratio $R^b \equiv P^b/P^A$, and some higher order terms that we do not compute cancel in the ratio.  The green shaded region is the size of the theoretical error, which we have calculated by estimating the size of the two-loop corrections that we have not included, using Eqs. (\ref{deltab1}) and (\ref{deltab2}), and the green dashed line is this error added in quadrature with a $1\%$ error for unknown systematics.  }  \label{wmap3_barfitratio}
\end{center}
\end{figure}

\begin{figure}[htb!] 
\centering
\begin{tabular}{cc}
\includegraphics[width=8cm]{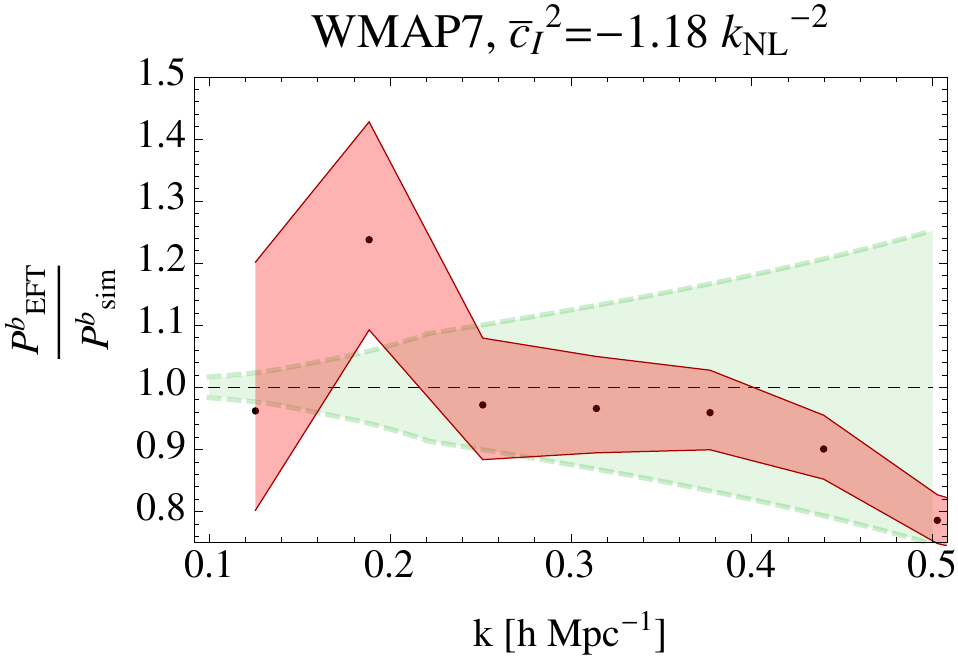} & \includegraphics[width=8.1cm]{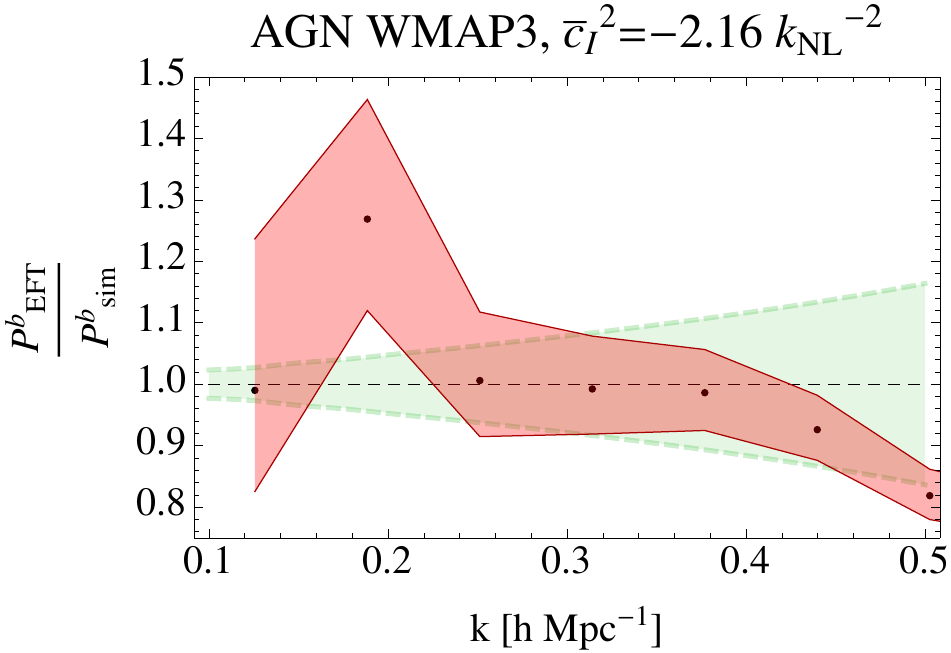}
\end{tabular}
\caption{  Here we compare the value of the baryon power spectrum $P^b$ calculated in the EFT with the given values of $\bar{c}_I^2$, to the simulation data.  This figure shows that the values of $\ci$ determined in the plots above, which use the ratio $R^b$ to minimize cosmic variance, are well within the cosmic variance of the raw data $P^b_{\rm sim}$.  Notice that the fits start to fail at $k \approx 0.4 \unitsk$.  This is expected within the EFT because it is where two-loop terms start to become important.  The green shaded region is the size of the theoretical error, which we have calculated by estimating the size of the two loop corrections that we have not included, in the same way as \eqn{estimate11}, and the green dashed line is this error added in quadrature with a $1\%$ error for unknown systematics.}  \label{pbratios}
\end{figure}

Again, we see that the EFT is able to describe a wide range of effects of the baryons not only on the dark matter, but also on the baryons themselves.  By fitting the parameters $\ca$ and $\ci$, the EFT at one-loop level correctly describes the effects of baryons at the $2\%$ level for $k \lesssim 0.4 \unitsk$, and at $k\lesssim 0.7\hinvMpc$ for the ratios of the power spectra.  All of the baryonic physics is encoded in the definite functional form $k^2 P_{11}$, and only two numerical coefficients are necessary to produce all these quantities.  

It is interesting to note the relatively small value of $\ci \sim 1 \,\knl^{-2}$ compared to the value of $\bar c_A^2\sim8\knl^{-2}$. The parameter $\ci$ is a combination of the differences of the speed of sounds for dark matter and baryons, plus the speed of sound induced on baryons from star formation. All of these differences vanish in the limit of no star formation, and so does $\ci$~\footnote{$\bar c_I^2$ is nonzero also because the initial conditions of baryons and dark matter are different. We check in Appendix~\ref{app:darkmattersim} that the difference induced by this is negligibly small.}. The numerical value that we find therefore seems to indicate that star formation processes have a smaller effect on baryons than gravitational non-linearities do.

Finally, we make a comment of the measurability of the quantities $\Delta \bar c_A^2$ and $\bar c_I^2$ from simulations. We can see that by using the ratios (\ref{eq:ra}) and (\ref{eq:ratiob}), which have extremely small cosmic variance, we can accurately measure  $\Delta \bar c_A^2$ and $\bar c_I^2$ from very small simulations of box size $L = 100 \, h^{-1} \rm Mpc$, which have large cosmic variance for power spectra, as can be seen from Fig.s~(\ref{determinecs}) and (\ref{pbratios}). This suggests that it is possible for baryonic simulations to focus on short distances and increase their accuracy.

%
%
 %
 %

\section{Conclusions}

We have argued that the effect of baryons at large distances in our universe can be accurately described by an effective field theory that consists of two fluid-like species, dark matter and baryons, interacting only gravitationally. What makes an effective treatment possible even for baryonic effects is that baryons do not move much due to star formation: the momentum transfer generated by star-formation physics is negligible. The effect of short distance non-linearities and star formation is encoded in two effective stress tensors, one for dark matter and one for baryons. The difference between dark matter and baryons, which comes from the different initial conditions and the star-formation physics, is encoded in the different sizes of the numerical coefficients that appear in the stress tensor.  This tells us that the {\it functional form} of the effect of baryons at large distances is predicted analytically, up to some unknown prefactors that need to be measured in observations (as we normally do when we measure the conductivity of a metal in a laboratory), or from small numerical simulations (as we do when we simulate the conductivity of the same metal on a powerful computer). The leading effect comes from a gravitationally and star-formation induced speed of sound, which affects the power spectrum in a form given by
\be\label{eq:concl}
\Delta P_b(k)\propto \left(\frac{k}{\knl}\right)^2 P_{11}^A(k)\ , \qquad \Delta P_c(k)\propto \left(\frac{k}{\knl}\right)^2 P_{11}^A(k) \ .
\ee
The different star formation models predict the different numerical prefactor to these terms.

In the EFTofLSS, which we extend in this paper to include baryonic effects, the clustering at large distances is organized in a perturbative expansion in powers of two parameters $\epsilon_{\delta<}$ and $\epsilon_{s>}$, representing the effect of tidal forces and short wavelength displacement, that become order one at the non-linear scale. This tells us that in the EFTofLSS each perturbative order is expected to perform better than the lower order, and it is supposed to agree with the numerical data up to a certain wavenumber that can be estimated before actually doing the calculations.

Naively, the EFTofLSS expands also in a parameter $\epsilon_{s<}$, which represents the effect of long wavelength displacements. This parameter is order one for the modes of interests, and therefore we resum all the effects proportional to $\epsilon_{s<}$, effectively not expanding in this parameter. This is achieved by extending the IR-resummation developed in~\cite{Senatore:2014via} to the case of baryons. Interesting, this generalization allows us to develop a formalism that allows us to consistently compute the effect of the relative motion of baryons at high redshift, an effect first noticed in~\cite{Tseliakhovich:2010bj}. 

We have checked our claims by comparing the predictions of the EFTofLSS against several quantities obtained in numerical simulations that include different modeling of baryonic effects. We have found that different star formation models predict an effect at large distances whose functional form agrees with (\ref{eq:concl}), and an overall prefactor that changes according to the different models. Inspection of the actual numerical size of the first terms in the perturbative series have taught us that the non-linear scale for baryons is not appreciably different than the one of dark matter, indicating that the effect of star formation on baryons is not larger than the one due to gravity.  This has allowed us to predict when a calculation for a given quantity at a given order should stop matching the data, finding percent agreement up to relatively high wavenumber such as $k\simeq 0.3\hinvMpc$ or $k\simeq 0.6\hinvMpc$, depending on the theoretical error of the calculation.

We find our results interesting for several reason. First, they provide a simple analytical understanding of the effect of baryonic physics on large scale structures. Second, they indicate that even if we were not to have available  reliable numerical simulations, the effect of baryons can be well accounted for in observations by directly fitting for the additional one parameter (or few parameters for higher order calculations). Third, and most importantly, they suggest that simulations do not need to provide power spectra of baryonic effects at large distances, as these can be computed with the EFTofLSS, but can instead focus on shorter distances, potentially increasing the accuracy of their predictions by a large amount~\footnote{In fact, by focusing on ratios between quantities computed with and without baryonic physics, we are able to measure the relevant coefficients of the stress tensor with great accuracy using small numerical simulations.}.

Though many additional computations and checks need to be done, our results are of great encouragement to the EFTofLSS program. This is a research program that aims at enabling us to make a more powerful use of large scale structure data. If it continues to be as successful as it has been so far from what we described in the introduction, the EFTofLSS suggests that the cosmological information available from next generation large scale structure experiments can be hugely superior to current expectations, therefore allowing us to largely increase our knowledge of the primordial conditions of the universe, beyond the remarkable level that the CMB has already allowed us to reach.

%
%
 %
 %

\subsubsection*{Acknowledgments}

We are greatly indebted to T,~Abel, J.~Schaye and M.~P.~van Daalen for providing us with the data of their simulations. We thank T.~Abel, S.~Allen, S.~Foreman H. Peiris, R.~Sheth, D.~Spergel, R.~Teyssier, and M.~Zaldarriaga for conversations, S.~Foreman for sharing some of his Mathematica codes, and T.~Abel, S.~Allen, N.~Arkani-Hamed, P.~Creminelli, L.~Dixon, H. Peiris, J.~Polchinski, R.~Rattazzi, N.~Seiberg, R.~Sheth, E.~Silverstein, D.~Spergel, R.~Teyssier, R.~Wechsler, M.~Wise and M.~Zaldarriaga for encouragement. M.~L.~ acknowledges support from the NSF through the GRF program. A.~P. is supported by the Gabilan Stanford Graduate Fellowship.
L.S.~is supported by by DOE Early Career Award DE-FG02-12ER41854 and the National Science Foundation under PHY-1068380.
This research was supported by the Munich Institute for Astro- and Particle Physics (MIAPP) of the DFG cluster of excellence ``Origin and Structure of the Universe''.

%
%
 %
 %

\begin{appendix}

%
%
%
%

\section{Simulation of Two Dark Matter Fluids\label{app:darkmattersim}}

In this section we compare our results with the results obtained in \cite{Angulo:2013qp}, which simulated two  dark matter fluids, with different initial conditions and abundances. These are chosen to mimic the ones of baryons and dark matter in the current universe.  Although there is no baryonic physics included in this simulation, they non-linearly evolves the different initial conditions, and give therefore a measure of the importance of these initial conditions.  The cosmological parameters used were $\Omega_m = 0.276$, $\Omega_\Lambda = 0.724 $, $\Omega_b = 0.045$ , $h = 0.703$, $\sigma_8 = 0.811$, and $n_s = 0.961$.  

\begin{figure}[htb!] 
\begin{center}
\includegraphics[scale=1.3]{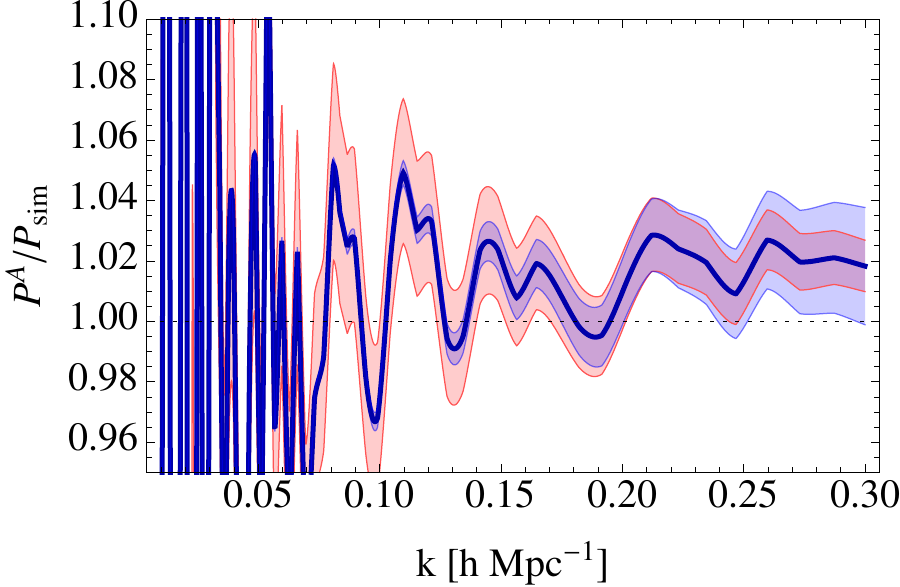}
\caption{   This plot compares our prediction  for the adiabatic power spectrum $P^A$ to the non-linear data from \cite{Angulo:2013qp}, and is used to determine our parameter $\ca$.  Here, the dark blue line has $\ca = 0.28 \unitscss$, and the blue shaded region has $\ca =  0.28 \pm w_b/(2\pi) \unitscss$.  The red shaded region is the cosmic variance of the simulation data due to a box size of $L = 1000\, h^{-1} \rm Mpc$.}\label{total_compare}
\end{center}
\end{figure}

The two parameters that we would like to match with the simulation are $\ca$ and $\bar{c}_I^2$. The total matter power spectrum can be used to determine $\ca$, and the difference between the power spectra of baryons and dark matter can be used to determine $\bar{c}^2_I$. In Fig.~\ref{total_compare} we compare our calculation of the total matter power spectrum, which in our notation is the IR-resummed adiabatic power spectrum~$P^{A}$ obtained using \eqn{powerspectra}, to the results of the simulation. Since $\bar{c}^2_I$ does not appear in the total matter power spectrum, we can use this to determine $\ca$. We find $\ca\sim 0.28 \unitscss$.

\begin{figure}[htb!] 
\begin{center}
\includegraphics[scale=1.2]{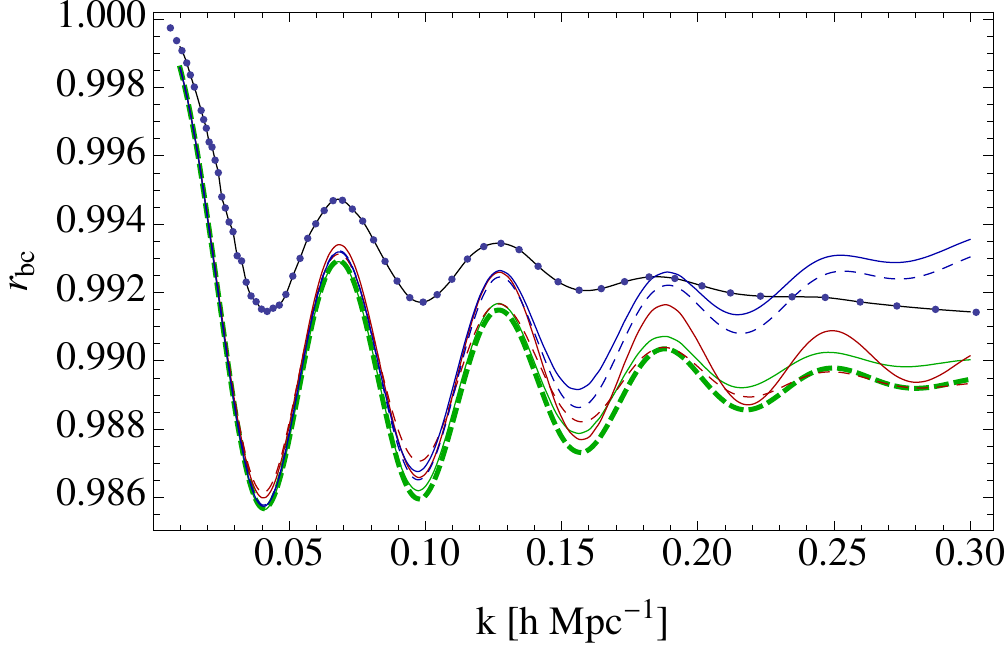}
\caption{  Comparison of IR-resummed and non-IR-resummed results for $r_{bc}=P^b/P^c$, the ratio of the baryon power spectrum to the dark matter power spectrum, to nonlinear simulation.  The data points are directly from the simulation data, and the black curve is the interpolation of those points.  The data of the \emph{ratio} of $P^b$ and $P^c$ have much less cosmic variance than the absolute power spectra.  In red are the 0-loop ratios, in green are the EFT 1-loop ratios with $\bar{c}_I^2 =-5\times 10^{-3} ( h \rm Mpc^{-1} )^{-2}$, and in blue are the EFT 1-loop ratios with  $\bar{c}_I^2 =0$. The dotted lines are the IR-resummed quantities and the solid lines are the non-resummed quantities.  All of the approximations are within $0.6\%$ of the simulation. Since we have already neglected $1\%$ effects such as the decaying modes and the isocurvature modes in loops, the differences between the curves in these plots are not significant.} 
\label{rbc} 
\end{center}
\end{figure}

Next, we would like to determine $\bar{c}^2_I$, which can be done by fitting the expressions in \eqn{powerspectra} to the nonlinear data. Just as we did in the Section \ref{hiratasection}, we will implement the IR-resummation separately on the baryon and dark matter power spectra. Fig.~\ref{rbc} shows the result of this computation in the ratio of the baryon power spectrum to the dark matter power spectrum compared to the numerical simulation. The best fit to the numerical simulation is $\bar{c}_I^2 =-5\times 10^{-3} ( h \rm Mpc^{-1} )^{-2}$, but to the precision of our calculation the results are also consistent with $\bar{c}_I^2 =0$. We expect only a very small isocurvature speed of sound because the simulation was done without including star formation processes; the only difference in the evolution of baryons and dark matter is a result of their different initial conditions. This causes a small deviation from the behavior of a single fluid, which is encoded in $\bar{c}_I^2$.

\end{appendix}

 \begingroup\raggedright\endgroup

 \end{document}